\documentclass[12pt]{article}      
\usepackage{natbib,epsfig,multirow,setspace}

\AtEndOfClass{\RequirePackage{times}}
\usepackage{mathtools}
\usepackage{amsthm,amsmath,amssymb,colonequals,multirow,graphicx,epstopdf}
\usepackage{caption}
\usepackage{subcaption}
\usepackage{color}
\usepackage{mwe}
\newcommand{\diag}{\,\mbox{diag}}
\newcommand{\tlog}{\,\mbox{log}}

\newcommand{\noisev}{\mathbf\Psi}

\newcommand{\ident}{\mathbf{I}}
\newcommand{\vecY}{\mathbf{Y}}
\newcommand{\vecy}{\mathbf{y}}
\newcommand{\vecx}{\mathbf{x}}
\newcommand{\vecc}{\mathbf{c}}
\newcommand{\vecC}{\mathbf{C}}

\newcommand{\vecV}{\mathbf{V}}

\newcommand{\vecZ}{\mathbf{Z}}

\newcommand{\Ew}{\mathbb{E}\left[ W_{ik} \mid \vecx_i, \vecy_i, c_{ik} = 1 \right]}
\newcommand{\Ewinv }{\mathbb{E}\left[1/W_{ik} \mid \vecx_i, \vecy_i, c_{ik} = 1 \right]}
\newcommand{\Elogw}{\mathbb{E}\left[ \log{W_{ik}} \mid \vecx_i, \vecy_i, c_{ik} = 1 \right]}
\newcommand{\Eeta}{\mathbb{E}\left[ \veceta_i \mid \vecy_i,\vecx_i,c_{ik}=1 \right]}
\newcommand{\Eetawinv}{\mathbb{E}\left[(1/W_{ik})\veceta_i \mid \vecy_i,\vecx_i,c_{ik}=1 \right]}
\newcommand{\Eetaetawinv}{\mathbb{E}[(1/W_{ik})\veceta_i \veceta_i^{'}\mid \vecy_i,\vecx_i,c_{ik}=1]}

\newcommand{\varthet}{\mbox{\boldmath$\vartheta$}}

\newcommand{\veceta}{\mbox{\boldmath$\eta$}}
\newcommand{\vecTheta}{\mbox{\boldmath$\Theta$}}
\newcommand{\vecmu}{\mbox{\boldmath$\mu$}}
\newcommand{\vecbeta}{\mbox{\boldmath$\beta$}}
\newcommand{\vecalpha}{\mbox{\boldmath$\alpha$}}
\newcommand{\veczeta}{\mbox{\boldmath$\zeta$}}
\newcommand{\vecepsilon}{\mbox{\boldmath$\epsilon$}}
\newcommand{\vecgamma}{\mbox{\boldmath$\gamma$}}
\newcommand{\vecpi}{\mbox{\boldmath$\pi$}}
\newcommand{\vecomega}{\mbox{\boldmath$\omega$}}
\newcommand{\veclambda}{\mbox{\boldmath$\lambda$}}
\newcommand{\mSigma}{\mbox{\boldmath$\Sigma$}}
\newcommand{\matsig}{\mSigma}
\newcommand{\mLambda}{\mbox{\boldmath$\Lambda$}}
\newcommand{\mGamma}{\mbox{\boldmath$\Gamma$}}
\newcommand{\lprod}{\prod_{i=1}^n}
\newcommand{\lsum}{\sum_{i=1}^n}

\newcommand{\ksum}{\sum_{k=1}^K}
\newcommand{\kprod}{\prod_{k=1}^K}

\DeclareMathOperator{\E}{\mathbb{E}}

\newcommand\inv[1]{#1\raisebox{1.15ex}{$\scriptscriptstyle-\!1$}}

\begin{document}

\title{Extending Growth Mixture Models Using Continuous Non-Elliptical Distributions}

\author{\ \ Yuhong Wei\thanks{Department of Mathematics \& Statistics, McMaster University, Hamilton, ON, Canada.} \qquad Yang Tang$^*$\qquad Emilie Shireman\thanks{Department of Psychological Sciences, University of Missouri, Columbia, MO, U.S.}\\[+4pt] Paul~D.~McNicholas$^*$ \quad Douglas L.\ Steinley$^{\dagger}$}
\date{}

\maketitle

%%%%%%Abstract%%%%%%%
\begin{abstract}
Growth mixture models (GMMs) incorporate both conventional random effects growth modeling and latent trajectory classes as in finite mixture modeling; therefore, they offer a way to handle the unobserved heterogeneity between subjects in their development. GMMs with Gaussian random effects dominate the literature. When the data are asymmetric and/or have heavier tails, more than one latent class is required to capture the observed variable distribution. Therefore, a GMM with continuous non-elliptical distributions is proposed to capture skewness and heavier tails in the data set. Specifically, multivariate skew-t distributions and generalized hyperbolic distributions are introduced to extend GMMs. When extending GMMs, four statistical models are considered with differing distributions of measurement errors and random effects. The mathematical development of GMMs with non-elliptical distributions relies on their expression as normal variance-mean mixtures and the resultant relationship with the generalized inverse Gaussian distribution. Parameter estimation is outlined within the expectation-maximization framework before the performance of our GMMs with non-elliptical distributions is illustrated on simulated and real data.
\end{abstract}

%%%%%%Introduction%%%%%%
\section{Introduction}\label{sec:intro}

Many longitudinal studies focus on investigating how individuals change over time with respect to a characteristic that is measured repeatedly for each participant. Conventional random effect growth modeling has provided a number of tools for modeling intra-individual change and inter-individual differences in change \citep[e.g.,][]{laird82,bryk87,bryk92,mcardle87,singer03}, where within-class changes are described as a function of time and between-class changes are described by random effects and coefficients. Conventional random effects models provide a basis for formulating growth mixture models (GMMs) for longitudinal data. In the latent variable framework, a quadratic random effects growth model with a continuous outcome $y_{it}$ for individual $i$ at time $t$, and time-invariant covariates $x_i$, is specified according to
\begin{align}
\label{eqn:quadgrowth1}
y_{it} &= \eta_{0i}+\eta_{1i}(a_t - a_0)+\eta_{2i}(a_t - a_0)^2+\epsilon_{it},\\
\eta_{0i} &= \alpha_0 + \vecgamma_0'\vecx_i +\zeta_{0i},\nonumber\\
\label{eqn:quadgrowth2}
\eta_{1i} &= \alpha_1 + \vecgamma_1'\vecx_i +\zeta_{1i},\\
\eta_{2i} &= \alpha_2 + \vecgamma_2'\vecx_i +\zeta_{2i},\nonumber
\end{align}
where $a_t$ are time scores ($t=1,2,\ldots,T$) centred at $a_0$, $\eta_{0i}$ is the random intercept, $\eta_{1i}$ is the random linear slope, $\eta_{2i}$ is the quadratic growth rate, and $\vecepsilon_i$ and $\veczeta_i$ are normally distributed residuals, i.e., $\vecepsilon_i \sim \mathcal{N}(\mathbf{0},\vecTheta)$ and $\veczeta_i \sim \mathcal{N}(\mathbf{0},\noisev)$. Formally, $\eta_{0i}$, $\eta_{1i}$, and $\eta_{2i}$ are continuous latent variables (called the growth factors) representing the growth patterns, the $\alpha_k$ are the mean parameters for the growth factors if there are no covariates $\vecx_i$, and the $\vecgamma_k$ are the effects of covariates $\vecx_i$ on the growth factors. In conventional growth modeling applications, the individual growth parameters (e.g., individual intercept and slope factors) are usually assumed to be identically distributed, i.e., drawn from a single homogeneous population. However, we are often interested in and deal with samples from multiple populations and, in most cases, the class memberships are either unknown or unobserved.

Simultaneous modeling of change over time and unobserved multiple populations (heterogeneity) in the data can be accommodated using GMMs and latent class growth analysis (LCGA), wherein parameters describing growth patterns are estimated and each individual's most likely class membership is obtained via maximum \emph{a~posteriori} (MAP) probabilities. GMMs were introduced by \citet{verbeke96} and then extended by \citet{muthen99}, \citet{muthen04}, and \citet{muthen08}. For convenience, define $c_{ik}$ so that $c_{ik} = 1$ if individual $i$ falls in class $k$ and $c_{ik} = 0$ otherwise. The quadratic growth model in \eqref{eqn:quadgrowth1} and~\eqref{eqn:quadgrowth2} can be extended to a simple GMM in class $k$ ($k=1,2,\ldots,K$) via
\begin{align}
\label{eqn:quadmixturegrowth1}
Y_{it} \mid {c_{ik}=1}  &= \eta_{0i}+\eta_{1i}(a_t - a_0)+\eta_{2i}(a_t - a_0)^2+\epsilon_{it},\\
\eta_{0i}\mid {c_{ik}=1} &= \alpha_{0k} + \vecgamma_0'\vecx_i +\zeta_{0i},\nonumber\\
\label{eqn:quadmixturegrowth2}
\eta_{1i}\mid {c_{ik}=1} &= \alpha_{1k} + \vecgamma_1'\vecx_i +\zeta_{1i},\\
\eta_{2i} \mid {c_{ik}=1} &= \alpha_{2k} + \vecgamma_2'\vecx_i +\zeta_{2i},\nonumber
\end{align}
where $\vecalpha_k = (\alpha_{0k},\alpha_{1k},\alpha_{2k})$ parameters vary across classes to capture different trajectories; the parameters $\vecgamma_0, \vecgamma_1$, and $\vecgamma_2$ remain the same across classes but this could be relaxed to allow variation across classes with respect to how a covariate affects the growth factors; and $\vecepsilon_i$ and $\veczeta_i$ are still normally distributed residuals but with class-specific covariance matrices, i.e., $\vecepsilon_i \sim \mathcal{N}(\mathbf{0},\vecTheta_k)$ and $\veczeta_i \sim \mathcal{N}(\mathbf{0},\noisev_k)$. The latent class growth analysis (LCGA) developed by Nagin and Land \citep{nagin93,nagin99,nagin05} can be thought of as one special case of GMMs in the sense that it assumes zero within-class growth factor variances, i.e., $\noisev_k=\mathbf{0}$ for $k=1,2,\ldots,K$.

One common fundamental assumption for GMMs is that model errors are normally distributed. However, simulation studies in \citet{bauer03} show that, when the data are drawn from a single non-Gaussian distribution, a two-class Gaussian GMM is preferred when fitting the data. In such cases, the within-class parameter estimates become uninterpretable because there are too many groups. \citet{muthen15} give an example of strongly non-normal outcomes, i.e., body mass index (BMI) development over age, and show that more than one latent class is required to capture the observed variable distribution --- interpreting mixture components as subpopulations will lead to overestimation of the number of subpopulations. In general, non-elliptical distributions are used in multivariate analysis for the study of asymmetric data; such distributions can have a concentration parameter to account for heavy tailed data. Intuitively, in the two-dimensional case, the joint distribution forms a non-elliptical shape in the iso-density plot. Relaxing the normality assumption for asymmetry and skewness, \citet{muthen15} develop a GMM with a normally distributed model error and ``classical" multivariate skew-t (cMST) random effects, i.e., $\vecepsilon_i \sim \mathcal{N}(\mathbf{0},\vecTheta_k)$ and $\veczeta_i \sim \text{MST}(\mathbf{0},\noisev_k)$.  An alternative specification of GMMs (called nonlinear mixed effect mixture models) was developed by \citet{lu14} within a Bayesian framework, wherein $\vecepsilon_i$ is assumed to follow a cMST while $\veczeta_i$ remains normally distributed. Note that the ``classical" formulation of the multivariate skew-t distribution is that given by \cite{azzalini96} and \cite{azzalini99}. In this paper, we outline a more general extension of GMMs to the generalized hyperbolic distribution while also considering the formulation of the multivariate skew-t distribution that arises as its special and limiting case (Section~\ref{sec:back}). The advantage of the generalized hyperbolic distribution is its flexibility. Many other well-known distributions are special or limiting cases of the generalized hyperbolic distribution; please refer to \citet[][]{mcneil05} for details on a variety of limiting cases of the generalized hyperbolic distribution.

The remainder of this article is laid out as follows. In Section~\ref{sec:back}, we go through some important background material on the generalized hyperbolic distribution as well as a special and limiting case that gives a formulation of the multivariate skew-t distribution. In Section~\ref{sec:method}, we outline the extension of GMMs to generalized hyperbolic and multivariate skew-t distributions, respectively. Section~\ref{sec:parameter} presents an expectation-maximization (EM) algorithm for obtaining maximum likelihood estimates of model parameters. Then, our approach is illustrated on simulated and real data (Section~\ref{sec:illustration}). The paper concludes with some discussion and suggestions for future work (Section~\ref{sec:discussion}).

%%%%%%Background%%%%%%%%
\section{Background}\label{sec:back}

%%%%%%%%%%
%GHD
%%%%%%%%%%
\subsection{Generalized hyperbolic distribution}\label{sec:backghyp}
A multivariate generalized hyperbolic distribution arises from a multivariate mean-variance mixture where the weight function $h(w\mid\omega,\eta,\lambda)$ is the density of a generalized inverse Gaussian (GIG) distribution. The density of the GIG distribution is given by 
\begin{equation}
\label{eqn:gig}
h(w\mid\omega,\eta,\lambda) = \frac{(w/\eta)^{\lambda-1}}{2\eta K_\lambda(\omega)}\text{exp}\left\{-\frac{\omega}{2}\left(\frac{w}{\eta}+\frac{\eta}{w}\right)\right\}
\end{equation}
for $w>0$, where $\eta>0$ is a scale parameter, $\omega>0$ is a concentration parameter, $K_{\lambda}$ denotes the modified Bessel function of the third kind with index $\lambda$, and $\lambda$ characterizes certain subclasses and considerably influences the size of tail weights. Write $W \sim \mathcal{I} (\omega,\eta, \lambda)$ to denote that the random variable $W$ has the density in \eqref{eqn:gig}. The GIG distribution has some attractive features, including tractable expected values. Consider $W \sim \mathcal{I} (\omega,\eta, \lambda)$, then the following expected values hold:
\begin{align}
\label{eqn:expectedvalue}
\E [W] &= \eta \frac{K_{\lambda+1}(\omega)}{K_{\lambda}(\omega)},\nonumber\\
\E [1/W] &= \frac{1}{\eta}\frac{K_{\lambda+1}(\omega)}{K_\lambda(\omega)}-\frac{2\lambda}{\omega\eta},\\
\E [\log W] &= \log\eta+\frac{1}{K_\lambda(\omega)}\frac{\partial }{\partial \lambda}K_\lambda(\omega).\nonumber
\end{align} Extensive details on GIG distribution can be found in \citet{Jorgensen82}. 

\cite{browne15} show that a $p$-dimensional generalized hyperbolic random variable $\vecY$ can be generated using the relationship
\begin{equation}
\label{eqn:ghstochastic}
\vecY = \vecmu + W \vecbeta + \sqrt{W} \vecZ,
\end{equation}
where $W \sim \mathcal{I} (\omega,1, \lambda)$, $\vecmu$ and $\vecalpha$ are $p$-vectors that play the role of location and skewness parameters, respectively, and $\vecZ\sim\mathcal{N}(\mathbf{0},\matsig)$. 
From \eqref{eqn:ghstochastic}, it follows that $\vecY \mid w \sim \mathcal{N}(\vecmu+w\vecbeta,w\matsig)$. Now, recalling that $W \sim \mathcal{I}(\omega,1,\lambda)$ and that the unconditional distribution of $\vecY$ is a generalized hyperbolic, Bayes' theorem gives 
\begin{equation}\label{eqn:elegant}
W\mid\vecy \sim \text{GIG}( \omega +\vecbeta^{'}\matsig^{-1}\vecbeta, \omega +\delta (\vecy,\vecmu\mid\matsig),\lambda-p/2 ).
\end{equation} %This elegant result will be used to extend GMMs to the generalized hyperbolic distribution.
%
%Note that the density in \eqref{eqn:gig} is sometimes a more meaningful reparameterization of the GIG distribution, denoted by $\text{GIG}(\psi,\chi,\lambda)$~\citep{good53,barndorff77,blaesild78,halgreen79,Jorgensen82}. The links between these two parameterizations of the GIG distribution are $\omega = \sqrt{\psi\chi}$ and $\eta=\sqrt{\chi/\psi}$. 
%Extensive details on the derivative of these parameterizations and the relationship in \eqref{eqn:ghstochastic} are given by \citet{browne15}. 
Under this parameterization, a $p$-dimensional multivariate generalized hyperbolic distribution has  density
\begin{equation}
\label{eqn:ghd}
f(\vecy \mid \varthet) = \left[\frac{\omega +\delta (\vecy,\vecmu\mid\matsig)}{\omega +\vecbeta^{'}\matsig^{-1}\vecbeta} \right]^{(\lambda-p/2)/2} \frac{K_{\lambda-p/2}\left(\sqrt{\left[ \omega +\vecbeta^{'}\matsig^{-1}\vecbeta\right]\left[\omega +\delta (\vecy,\vecmu\mid\matsig) \right]}\right)}{(2\pi)^{p/2}\mid\matsig\mid^{1/2}K_\lambda(\omega) \text{exp}\{-(\vecy-\vecmu)^{'}\matsig^{-1}\vecbeta\}},
\end{equation}
with index parameter $\lambda $, concentration parameter $\omega$, skewness parameter $\vecbeta$, mean vector $\vecmu$, and scale matrix $\matsig$. Here, $\delta (\vecy, \vecmu\mid\matsig)$ is the squared Mahalanobis distance between $\vecy$ and $\vecmu$, i.e., $\delta (\vecy, \vecmu\mid\matsig) = (\vecy - \vecmu)^{'}\matsig^{-1}(\vecy - \vecmu)$, $K_{\lambda-{p}/{2}}$ and $K_\lambda$ are modified Bessel functions of the third kind with indices $\lambda-{p}/{2}$ and $\lambda$, respectively, and $\varthet= (\lambda, \omega, \vecmu, \matsig,\vecbeta)$ denotes the model parameters. Herein, let $\vecY \sim \text{GHD}_p (\lambda, \omega, \vecmu, \matsig,\vecbeta)$ represent a $p$-dimensional generalized hyperbolic random variable $\vecY$ with density as per \eqref{eqn:ghd}. Note that the parameterization in \eqref{eqn:ghd} is one of several available for multivariate generalized hyperbolic distributions \citep[see][]{mcneil05, browne15}.

\begin{figure}[htb]
        \centering
      \includegraphics[width=0.975\textwidth]{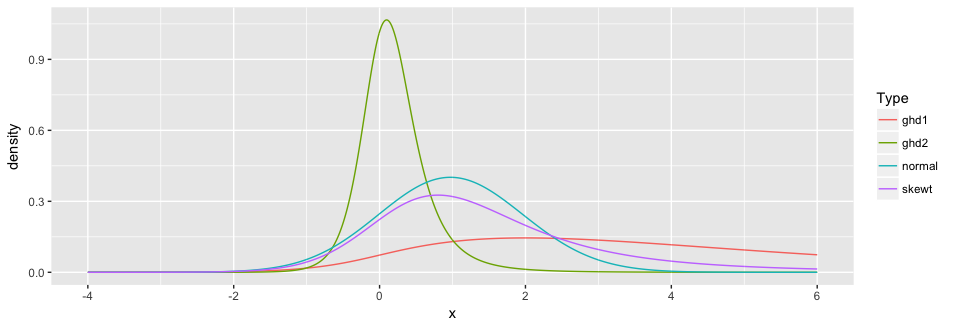}
      \caption{{\small Density plot for the Gaussian distribution (blue), skew-t distribution with $\nu=5$ (purple), and the generalized hyperbolic distribution with $\lambda=2$ (red) and $\lambda=-2$ (green), respectively. }}\label{fig:densityplot}
\end{figure}

There are a number of special and limiting cases that can be derived from the generalized hyperbolic distribution. However, the presence of the index parameter $\lambda$ enables a flexibility that is not found in its special and limiting cases. Figure~\ref{fig:densityplot} illustrates the Gaussian distribution as well as a skew-t distribution with $\nu=5$ degrees of freedom and the generalized hyperbolic distribution for two different values of $\lambda$; clearly, the different values of $\lambda$ lead to very different densities.

%%%%%%%%%%%%
%SkewT
%%%%%%%%%%%%
\subsection{Multivariate skew-t distribution}\label{sec:mvskewt}
Several alternative formulations of the multivariate skew-t distribution have appeared in the literature, e.g., \cite{azzalini96}, \cite{sahu03}, and \cite{arellano05}. Some recent discussion about some of these formulations is given by \cite{azzalini16}. \citet{muthen15} developed a GMM with the SDB version of the restricted multivariate skew-t distribution, i.e., the version of \cite{sahu03}. The formulation of the multivariate skew-t distribution used herein arises as a special and limiting case of the generalized hyperbolic distribution by setting $\lambda=-\nu/2$ and $\chi=\nu$ while also letting $\psi \to 0$. This formulation of the multivariate skew-t distribution has been used by \citet{murray14a} to develop a mixture of skew-t factor analyzer models and by \citet{murray14b} to develop a mixture of common skew-t factor analyzers.
 A $p$-dimensional skew-t random variable $\vecY$, with this formulation, has the density function 
 \begin{equation}
\label{eqn:skewt}
f(\vecy \mid \varthet) = \left[\frac{\nu+\delta (\vecy,\vecmu\mid\matsig)}{\vecbeta^{'}\matsig^{-1}\vecbeta} \right]^{(-\nu-p)/4} \frac{\nu^{\nu/2}K_{(-\nu-p)/2}(\sqrt{\left[ \vecbeta^{'}\matsig^{-1}\vecbeta\right]\left[\nu +\delta (\vecy,\vecmu\mid\matsig) \right]})}{(2\pi)^{p/2}\mid\matsig\mid^{1/2} \Gamma(\nu/2)^{\nu/2} \text{exp}\{-(\vecy-\vecmu)^{'}\matsig^{-1}\vecbeta\}},
\end{equation}
where $\vecmu$ is the location parameter, $\matsig$ is the scale parameter, $\vecbeta$ is the skew parameter, $\nu$ is the degree of freedom parameter, and $K_{(-\nu-p)/2}$ and $\delta (\vecy,\vecmu\mid\matsig)$ are as defined in \eqref{eqn:ghd}. We write $\vecY \sim \text{GST}(\vecmu,\matsig,\vecbeta,\nu)$ to denote that the random variable $\vecY$ follows the skew-t distribution with the density in \eqref{eqn:skewt}. Now, $\vecY \sim \text{GST}(\vecmu,\matsig,\vecbeta,\nu)$  can be obtained through the relationship in \eqref{eqn:ghstochastic} with $W \sim \text{IG}(\nu/2,\nu/2)$, where $\text{IG}(\cdot)$ denotes the inverse-gamma distribution. We have $\vecY \mid w \sim \mathcal{N}(\vecmu+w\vecbeta,w\matsig)$, and so, from Bayes's theorem, $W \mid \vecy \sim \text{GIG}( \vecbeta^{'}\matsig^{-1}\vecbeta, \nu +\delta (\vecy,\vecmu\mid\matsig),-(\nu+p)/2 )$. %Similarly, this nice characterization will facilitate extension of the GMMs to consider the multivariate skew-t distributions.

%%%%%%%%%%%%%%
%EM and it's convergence
%%%%%%%%%%%%%%
\subsection{The EM algorithm and its convergence criterion}
The EM algorithm \citep{dempster77} is an iterative algorithm for finding maximum likelihood estimates when data are incomplete or treated as such, and is widely used to estimate model parameters in the context of model-based clustering.~The E-step computes the expected value of the complete-data log-likelihood given the current model parameters, and the M-step maximizes this expected value with respect to the model parameters. After each E- and M-step, the log-likelihood is driven uphill, and the method iterates towards a maximum until some convergence criterion is satisified. 
Many variants of the EM algorithm have been proposed over the years, such as the expectation-conditional-maximization (ECM) algorithm \citep{meng93}, the alternating ECM (AECM) algorithm \citep{meng97}, and the Fisher-EM algorithm \citep{bouveyron12}. Herein, we will make use of the EM algorithm for parameter estimation and a stopping criterion based on the Aitken acceleration \citep{aitken26} is used to determine the convergence. The Aitken acceleration at iteration $s$ is
\begin{equation}
a^{(s)} = \frac{l^{(s+1)}-l^{(s)}}{l^{(s)}-l^{(s-1)}},
\end{equation}
where $l^{(s)}$ is the (observed) log-likelihood value at iteration $s$. This yields an asymptotic estimate of the log-likelihood at iteration $s+1$, given by $$l_{\infty}^{(s+1)} = l^{(s)} + \frac{1}{1-a^{(s)}}(l^{(s+1)}-l^{(s)})$$ \citep{bohning94, lindsay95}, and the EM algorithm is stopped when $l_{\infty}^{(s+1)}-l^{(s)} < \epsilon$, provided this difference is positive \citep{mcnicholas10}. Note that this criterion is at least as strict as the lack of progress criterion in the neighbourhood of a maximum. %, which identifies when the difference between successive log-likelihood values is less than some small $\epsilon$.

%%%%%%%%%%%%%%
%Model Selection
%%%%%%%%%%%%%%
\subsection{Model selection}
In model-based clustering, a penalized log-likelihood-based criterion is typically used to determine the ``best'' fitting model among a family of models. The most popular such criterion is the Bayesian information criterion \citep[BIC;][]{schwarz78}, which can be motivated as an approximation to a Bayes factor \citep[see][]{kass95,kasswass95}. The BIC is defined as 
\begin{equation}
\label{eqn:bic}
\text{BIC} = 2l(\hat{\varthet}) - \rho \log n,
\end{equation} 
where $\hat{\varthet}$ is the maximum likelihood estimate of model parameters $\varthet$, $l(\hat{\varthet})$ is the maximized log-likelihood, $\rho$ is the number of free parameters, and $n$ is the number of observations. Some theoretical support for use of the BIC in mixture model selection is given by \citet{leroux92} and \citet{keribin00}.

%%%%%%Methodology%%%%%%%%
\section{Methodology}\label{sec:method}

%%%%%%%%%%%%%%%
%conventional GMM
%%%%%%%%%%%%%%%
\subsection{Conventional GMM with Gaussian random effects}\label{sec:gmm}
Suppose a longitudinal study features $n$ subjects and $T$ time points or measurement occasions. For subject $i$ ($i = 1, \ldots, n$), let $\vecy_i$ be a $T \times 1$ vector $\vecy_i = (y_{i1}, y_{i2},\ldots,y_{iT})'$ where $y_{it}$ represents the outcome on occasion $t$ ($t = 1, \ldots, T$), let $\vecx_i = (x_{i1}, x_{i2}, \ldots, x_{im})^{'}$ be an $m \times 1$ vector of observed time-invariant covariates, let $\veceta_i$ be a $q \times 1$ vector containing $q$ continuous latent variables, and note that %let $\vecC_i$ be a $K \times 1$ vector consisting of $K$ indicator variables, i.e., 
$\vecC_i = (C_{i1},\ldots,C_{iK})'$ has a multinomial distribution, where $C_{ik} = 1$ if individual $i$ is in class $k$ and $C_{ik} = 0$ otherwise. The conventional GMM with Gaussian random effects can be represented using a hierarchical three-level formulation as follows.

At level 1 of the GMM, the continuous outcome variables $\vecY_1,\ldots,\vecY_n$ are related to the continuous latent variables $\veceta_1,\ldots,\veceta_n$ via
\begin{equation}
\label{eqn:yeta}
\vecY_i\mid (c_{ik}=1) = \mLambda_y\veceta_i+\vecepsilon_i
\end{equation} 
for $i=1,\ldots,n$, where $\vecepsilon_i$ is a $T \times 1$ vector of residuals or measurement errors that is assumed to follow a multivariate normal distribution, i.e., $\vecepsilon_i \sim \mathcal{N}(\mathbf{0}, \vecTheta_k)$, and $\mLambda_y$ is a $T \times q$ design matrix consisting of factor loadings with each column corresponding to specific aspects of change. The matrix $\mLambda_y$ and the vector $\veceta_i$ determine the growth trajectory of the model. For instance, when $q=3$,  $\veceta_i = (\eta_{0i}, \eta_{1i}, \eta_{2i})$, and $\mLambda_y$ is a $T \times 3$ matrix. Assuming $a_t$ are age-related time scores $(t = 1,2,\dots,T)$ centred at age $a_0$, the design matrix $\mLambda_y$ is given by
$$\mLambda_y = \begin{pmatrix}
				1 & a_1 - a_0 & (a_1 - a_0)^2\\
				1 & a_2 - a_0 & (a_2 - a_0)^2\\
				\vdots & \vdots&\vdots\\
				1 & a_{T-1} - a_0 & (a_{T-1} - a_0)^2
			\end{pmatrix}.$$

At level 2 of the GMM, the continuous latent variables $\veceta_i$ are related to the latent categorical variables $\vecc_i$ and to the observed time-invariant covariate vector $\vecx_i$ by the relation
\begin{equation}
\label{eqn:etacx}
\veceta_i\mid (c_{ik}=1) = \vecalpha_k + \mGamma_{k}  \vecx_i +\veczeta_i,
\end{equation}
where $\vecalpha_k$ $(k=1,\ldots,K)$ denotes the intercept parameter for class $k$, $\veczeta_i$ is a $q$-dimensional vector of residuals assumed to follow a multivariate normal distribution $\veczeta_i \sim \mathcal{N}(\mathbf{0}, \noisev_k)$, and $\mGamma_{k}$ is a $q\times m$ parameter matrix representing the effect of $\vecx_i$ on the latent continuous variables $\veceta_i$ and assumed to be different among classes. Note that the level 2 errors $\veczeta_i$ are uncorrelated with the measurement errors $\vecepsilon_i$. We may allow for class-specific effects $\mGamma_{k}$ in \eqref{eqn:etacx} that are equal across classes. 

By combining the first two levels of the GMM, we have
\begin{equation}
\label{eqn:gmm}
p (\vecy_i \mid \vecx_i) = \sum_{k=1}^{K} \pi_k \phi(\vecy_i ; \vecmu_k, \matsig_k),
\end{equation}
where $\pi_k = \text{Pr}(C_{ik} = 1)$ is the $k$th class probability or mixing proportion satisfying $\pi_k \in (0, 1]$ and $\sum_{k=1}^{K}\pi_k=1$, and $\phi(\cdot; \vecmu_k, \matsig_k)$ is a multivariate Gaussian density with mean $\vecmu_k = \mLambda_y(\vecalpha_k + \mGamma_k\vecx_i)$ and covariance matrix $\matsig_k = \mLambda_y\noisev_k\mLambda_y^{'}+\vecTheta_k$. Notice that the GMM in \eqref{eqn:gmm} assumes that class probability $\pi_k$ is constant for each class.

At level 3 of the GMM, we assume that the class probabilities are not constant for each class, but depend on the observed covariates. In other words, we want to know how $\pi_k$ is related to an individual's background variables, e.g., gender and income. At this level, the categorical latent variables $\vecC_{i}$ represent membership of mixture components that are related to $\vecx_i$ through a multinomial logit regression for unordered categorical responses. Define $\pi_{ik} = \text{Pr}(C_{ik}=1\mid\vecx_i)$, i.e., the probability that subject $i$ falls into the $k$th class depending on the covariates $\vecx_i$. Let $\vecpi_i=(\pi_{i1},\pi_{i2},\ldots,\pi_{iK})'$ and $$\text{logit}(\vecpi_i) = \left(\tlog\left(\frac{\pi_{i1}}{\pi_{iK}}\right),\tlog\left(\frac{\pi_{i2}}{\pi_{iK}}\right),\dots,\tlog\left(\frac{\pi_{i K-1}}{\pi_{iK}}\right)\right)'.$$ Then,
\begin{equation}
\label{eqn:cx}
\text{logit}(\vecpi_i) = \vecalpha_c + \mGamma_c \vecx_i,
\end{equation}
where $\vecalpha_c$ is a $(K-1)$-vector of parameters and $\mGamma_c$ is a $(K-1) \times q$ parameter matrix.
By combining these three levels of the GMM, we have
\begin{equation}
\label{eqn:egmm}
p (\vecy_i \mid \vecx_i) = \sum_{k=1}^{K} \pi_{ik} \phi(\vecy_i ; \vecmu_k, \matsig_k),
\end{equation}
where $\phi(\cdot; \vecmu_k, \matsig_k)$ is defined as in \eqref{eqn:gmm}. 
Note that the right hand side of \eqref{eqn:egmm} is not a finite mixture model because the class probabilities are not constant with resect to $i$.

%%%%%%%%%%%%%%%
%GHD-GMM
%%%%%%%%%%%%%%%
\subsection{GMM with generalized hyperbolic random effects}\label{sec:ghypgmm}
The conventional GMM assumes that the residuals $\vecepsilon_i$ and $\veczeta_i$ have multivariate Gaussian distribution with zero means and within-class covariance matrices, respectively. We are interested in constructing a GMM with generalized hyperbolic distribution model errors, denoted by GHD-GMM. The generalized hyperbolic distribution can be represented as a normal mean-variance mixture, where the mixing weight has a GIG distribution (see Section~\ref{sec:backghyp}). 
To this end, we introduce a latent continuous variable $W_{ik}$ with $W_{ik}\mid c_{ik}=1\sim \mathcal{I} (\omega_k, 1, \lambda_k)$. Accordingly, conditional on $c_{ik}$ and $w_{ik}$, we assume that model errors $\vecepsilon_i$ and $\veczeta_i$ are non-centered Gaussian error terms with distinct covariance matrices:
\begin{align}
\label{eqn:epsilon}
\vecepsilon_i \mid w_{ik},c_{ik}&=1 \sim \mathcal{N}(w_{ik}\vecbeta_{yk},w_{ik}\vecTheta_k), \\
\label{eqn:zeta}
\veczeta_i \mid w_{ik},c_{ik}&=1 \sim \mathcal{N}(w_{ik}\vecbeta_{\eta k},w_{ik}\noisev_k),
\end{align}
where $\vecTheta_k$ is the diagonal covariance matrix for $\vecepsilon_i$, and $\noisev_k$ is the covariance matrix for $\veczeta_i$. The $T$-dimensional vector $\vecbeta_{yk}$ is a vector of skewness parameters, which we refer to as the skewness parameter for the measurement errors. The $q$-dimensional vector $\vecbeta_{\eta k}$ is the vector of skewness parameters for the continuous latent variables $\veceta_i$. Then, based on \eqref{eqn:etacx} and \eqref{eqn:epsilon}, the observed random variables $\vecY_i$, conditional on $\veceta_i$, $c_{ik}$, and $w_{ik}$,  follow a conditional Gaussian distribution of the form
\begin{equation}
\label{eqn:observedyeta}
\vecY_i \mid\veceta_i,w_{ik},c_{ik}=1 \sim \mathcal{N}(\mLambda_y \veceta_i + w_{ik} \vecbeta_{yk}, w_{ik}\vecTheta_k ).
\end{equation}
Based on \eqref{eqn:etacx} and \eqref{eqn:zeta}, %the continuous latent variables $\veceta$ given $\vecx$, $\vecc$, and $w$, are normally distributed and has the density function
\begin{equation}
\label{eqn:eta}
\veceta_i \mid\vecx_i,w_{ik},c_{ik}=1 \sim \mathcal{N}(\vecalpha_k+\mGamma_{k} \vecx_i + w_{ik}\vecbeta_{\eta k}, w_{ik}\noisev_k)
\end{equation}
and, from the preceding equations, we have the conditional distribution
\begin{equation}
\label{eqn:observedy}
\vecY_i \mid \vecx_i, w_{ik}, c_{ik} = 1 \sim \mathcal{N} (\vecmu_k + w_{ik}(\mLambda_y\vecbeta_{\eta k} + \vecbeta_{yk}), w_{ik}\matsig_k),
\end{equation}
where $\vecmu_k=\mLambda_y(\vecalpha_k+\mGamma_{k}\vecx_i)$ and $\matsig_k=\mLambda_y\noisev_k\mLambda_y^{'}+\vecTheta_k$. From \eqref{eqn:elegant}, we obtain the conditional distributions
\begin{align}
\veceta_i\mid\vecx_i,c_{ik} = 1 &\sim \text{GHD}_q(\lambda_k,\omega_k,\vecalpha_k+\mGamma_{k}\vecx_i,\noisev_k,\vecbeta_{\eta k}),\\
\vecY_i\mid\vecx_i,c_{ik} = 1 &\sim \text{GHD}_T(\lambda_k,\omega_k,\vecmu_k,\matsig_k,\mLambda_y\vecbeta_{\eta k} + \vecbeta_{yk}).
\end{align}

By combining the preceding setup and level 3 of the GMM from Section~\ref{sec:gmm}, we arrive at a GMM with density %generalized hyperbolic distributions:
\begin{equation}
\label{eqn:ghdgmm}
p (\vecy_i \mid \vecx_i) = \sum_{k=1}^{K} \pi_{ik} f_{\text{GHD}_T}(\vecy_i ; \lambda_k, \omega_k, \vecmu_k, \matsig_k, \mLambda_y\vecbeta_{\eta k} + \vecbeta_{yk}),
\end{equation}where $f_{\text{GHD}_T}(\cdot)$ is the density of a $T$-dimensional random variable following a generalized hyperbolic distribution. Note that the overall skewness for $\vecY_i$ is $\mLambda_y\vecbeta_{\eta k} +\vecbeta_{yk}$. Note also that, within this setup, the dependent observed variable $\vecY_i$, the latent growth factors $\veceta_i$, and residual variables $\vecepsilon_i$ and $\veczeta_i$ all have generalized hyperbolic distributions. Note that the distribution of the covariates $\vecx_i$ is not modelled; please refer to \citet{muthen15} for detailed explanations.

%%%%%%%%%%%%%
%%GSTGMM
%%%%%%%%%%%%
\subsection{GMM with multivariate skew-t random effects}\label{sec:skewtgmm}

In this section, we are interested in extending the conventional GMM to have multivariate skew-t distribution model errors, denoted by GST-GMM. As in the case for the generalized hyperbolic distribution, the formulation of the multivariate skew-t distribution we use has a convenient representation as a normal mean-variance mixture; this time, the weight has an inverse-gamma distribution (see Section~\ref{sec:mvskewt}). In analogous fashion to the GHD-GMM, a latent continuous random variable $W_{ik}$ is first introduced, where $W_{ik} \mid c_{ik} = 1 \sim \text{IG}(\nu_k/2,\nu_k/2)$. Accordingly, we assume that $\vecepsilon_i$ and $\veczeta_i$ are non-centered Gaussian error terms with their own covariance matrices as in \eqref{eqn:epsilon} and \eqref{eqn:zeta}, and $\vecy_i$ and $\veceta_i$ are conditionally normally distributed as in \eqref{eqn:observedyeta} and \eqref{eqn:eta}. Form this characterization of the multivariate skew-t distribution, the following conditional distributions are obtained:
\begin{align}
\veceta_i\mid\vecx_i,c_{ik} = 1 &\sim \text{GST}_q(\vecalpha_k+\mGamma_{k}\vecx_i,\noisev_k,\vecbeta_{\eta k}, \nu_k),\\
\vecY_i\mid\vecx_i,c_{ik} = 1 &\sim \text{GST}_T(\vecmu_k,\matsig_k,\vecbeta_{yk} + \mLambda_y\vecbeta_{\eta k}, \nu_k),
\end{align}
where $\vecmu_k$ and $\matsig_k$  are as described above and $\nu_k$ is a concentration parameter (i.e., the degrees of freedom). Similarly, we arrive at a GMM with a multivariate skew-t distribution
\begin{equation}
\label{eqn:gstgmm}
p(\vecy_i\mid\vecx_i) = \ksum \pi_{ik} f_{\text{GST}_T}(\vecy_i;\vecmu_k,\matsig_k,\vecbeta_{yk}+\mLambda_y\vecbeta_{\eta k},\nu_k).
\end{equation}
In this setup, the random variable $\vecY_i$, the latent growth factors $\veceta_i$, and the residual variables $\vecepsilon_i$ and $\veczeta_i$ all follow multivariate skew-t distributions.

\subsection{Comments on the GHD-GMM and GST-GMM}

Recalling that the overall skewness for $\vecY_i$ is $\mLambda_y\vecbeta_{\eta k} +\vecbeta_{yk}$, there are a total of $T+q$ skewness parameters in our GMM extensions. Hence, the skewness parameters $\vecbeta_{yk}$ and $\vecbeta_{\eta k}$ are subject to identifiability issues because no more than $T$ skewness parameters can be identified from $T$-dimensional $\vecY_i$. Therefore, two special formulations are considered in this paper. The first formulation is where $\vecbeta_{yk} = \mathbf{0}$. In this formulation, the residuals for $\vecY_i$ or the measurement errors are not skewed, i.e., $\vecepsilon_i \mid w_{ik},c_{ik} =1 \sim \mathcal{N}(\mathbf{0},w_{ik}\vecTheta_k)$, and all of the skewness in the data is assumed to come from the distribution of latent factors. The second special formulation is the case where $\vecbeta_{\eta k} = \mathbf{0}$. In this formulation, the residuals for the latent factors $\veceta_i$ are symmetric, i.e., $\veczeta_i \mid w_{ik},c_{ik} =1 \sim \mathcal{N}(\mathbf{0},w_{ik}\noisev_k)$. Accordingly, all of the skewness in the data is assumed to come from the residuals of $\vecY_i$ or the measurement errors. In practice, we would want as much of the skewness as possible in the observed data $\vecY_1,\ldots,\vecY_n$ to be explained through the latent factors. There appears to be no optimal strategy with respect to which skewness parameter to estimate. Accordingly, four statistical models, differing with respect to the distributions of measurement errors and random effects for the first two levels of the GMM, are employed and compared. These models are as follows:
\begin{itemize}
\item \textbf{Model I:} A model with independent multivariate generalized hyperbolic random effects and measurement errors while assuming all of the skewness in the data comes from the distribution of latent factors (i.e., GHD-GMM under $\vecbeta_{yk} = \mathbf{0}$).
\item \textbf{Model II:} A model with independent multivariate generalized hyperbolic random effects and measurement errors while assuming all of the skewness in the data comes from the residuals of $\vecY_i$ (i.e., GHD-GMM under $\vecbeta_{\eta k} = \mathbf{0}$).
\item \textbf{Model III:} A model with independent multivariate skew-t random effects and measurement errors while assuming all of the skewness in the data comes from the distribution of latent factors (i.e., GST-GMM under $\vecbeta_{yk} = \mathbf{0}$).
\item \textbf{Model IV:} A model with independent multivariate skew-t random effects and measurement errors while assuming all of the skewness in the data comes from the residuals of $\vecY_i$ (i.e., GST-GMM under $\vecbeta_{\eta k} = \mathbf{0}$).
\end{itemize}
Take Model I (i.e., GHD-GMM under $\vecbeta_{yk}=\mathbf{0}$) as an example. For different trajectory classes, the parameters $\lambda_k, \omega_k, \vecalpha_k, \vecbeta_{\eta k}, \vecTheta_k, \noisev_k$, and $\mGamma_{k}$ may be different across classes, or may be the same across the classes. By imposing constraints on all these parameters (different or the same across classes), we obtain a family of GHD-GMM models. In this paper, we only consider two models, one model assumes that the parameters $\lambda_k, \omega_k, \vecalpha_k, \vecbeta_{\eta k}, \vecTheta_k, \noisev_k$,~and $\mGamma_{k}$ are different across classes, we call this model the general model. The second model assumes that only the parameter $\vecbeta_{\eta k}$ is different across classes while all the other parameters are the same across classes, i.e., $\lambda_k = \lambda$,  $\omega_k = \omega$, $\vecalpha_{k} = \vecalpha_{c}$, $\vecTheta_k = \vecTheta$, $\noisev_k=\noisev$, and $\mGamma_k=\mGamma_{c}$ for $k = 1, 2, \ldots, K$; we call this model the most constrained model. 

To this end, eight parameterizations in Table~\ref{nparmodels} are considered. Models~II and~IV allow a more general representation of the class skewness parameters (i.e., $\vecbeta_{yk}$). However, in terms of model complexity, Models~II and~IV %(GHD-GMM under $\vecbeta_{\eta k}=\mathbf{0}$) 
have $K(T-q)$ more parameters than Models~I and~III. %(GHD-GMM under $\vecbeta_{ yk}=0$), 
Hence, Models~II and~IV need larger sample sizes as small class sizes can create problems, such as singularity of the covariance matrix and slow or non-convergence of the EM algorithm. In addition, Model~III is the most parsimonious and it may be useful when the number of classes $K$ is large.
\begin{table}[!ht]
\caption{Key characteristics and the associated number of free parameters for the general and constrained varieties of Models~I--IV.}
\label{nparmodels}
	\centering
	%\resizebox{15.5cm}{!}{
	\small{
	%\footnotesize{
	\begin{tabular*}{1.0\textwidth}{@{\extracolsep{\fill}}lrrrr}
	\hline
\multicolumn{2}{c}{Model}&Dist.&Skewness&Number of free parameters\\
\hline
\multirow{2}{*}{Model I}&General&GHD&Latent $\veceta_i$&$K T+3K-1+K q  [(q+1)/{2}+2+m]$\\
&Constrained&GHD&Latent $\veceta_i$ &$T+K+1+q [({q+1})/{2}+K+1+m]$\\
\multirow{2}{*}{Model II}&General&GHD&Observed $\vecY_i$&$2 T  K +3K-1+K  q  [({q+1})/{2}+1+m]$\\
&Constrained&GHD&Observed $\vecY_i$&$T+2K+1+q[({q+1})/{2}+K+m]$\\
\multirow{2}{*}{Model III}&General&Skew-t&Latent $\veceta_i$&$K T+2K-1+K q  [({q+1})/{2}+2+m]$\\
&Constrained&Skew-t&Latent $\veceta_i$&$T+K+q [({q+1})/{2}+K+1+m]$\\
\multirow{2}{*}{Model IV}&General&Skew-t&Observed $\vecY_i$&$2 T  K+2K-1+K q [({q+1})/{2}+1+m]$\\
&Constrained&Skew-t&Observed $\vecY_i$&$T+2K+q [({q+1})/{2}+K+m]$\\
\hline
\end{tabular*}}
\end{table}

\subsection{Parameter estimation}\label{sec:parameter}
To fit the models, we adopt the well-known EM algorithm. In our case, the missing data comprise the latent categorical variables $\vecc_1,\ldots,\vecc_n$, the latent growth factors $\veceta_1,\ldots,\veceta_n$, and the latent weight parameter $w_{ik}$. Therefore, the complete-data consist of the observed outcome data $\vecy_1,\ldots,\vecy_n$, the covariates $\vecx_1,\ldots,\vecx_n$ together with the $\vecc_{i}$, $\veceta_{i}$, and $w_{ik}$, and complete-data likelihood is given by
\begin{equation*}
\mathcal{L}_{\text{c}}(\varthet) = \lprod\kprod[\pi_{ik}\phi(\vecy_i\mid\mLambda_y\veceta_i,w_{ik}\vecTheta_k)\phi(\veceta_i\mid\vecalpha_k+\mGamma_{k}\vecx_i+w_{ik}\vecbeta_{\eta k},w_{ik}\noisev_k)h(w_{ik})]^{c_{ik}},
\end{equation*}
where $W_{ik}\sim \mathcal{I} (\omega_k,\lambda_k)$ for GHD-GMM and  $W_{ik}\sim \text{IG}(\nu_k/2,\nu_k/2)$ for GST-GMM.

In the E-step, we compute the expected value of the complete data log-likelihood, denoted $\mathcal{Q}$, conditional on the current model parameters. Then, in the M-step, we obtain the updated model parameters by maximizing $\mathcal{Q}$. Detailed parameter updates for Models~I, II and~III are outlined in Appendix~B.

%%%%%%%Illustration%%%%%%%%
\section{Illustrations}\label{sec:illustration}
\subsection{Performance assessment}\label{sec:assessment}
Although all of our illustrations are treated as genuine clustering analysis, i.e., no prior knowledge of labels is assumed, the true labels are known in each case and can be used to evaluate the performance of our GHD-GMM and GST-GMM models. We use misclassification rates (ERR) and the adjusted Rand index \citep[ARI;][]{hubert85} to assess classification performance. The ERR is simply the proportion of misclassified observations. The ARI indicates the pairwise agreement between true and predicted group memberships while also accounting for the fact that random classification would classify some observations correctly by chance. An ARI value of 1 indicates perfect classification, its expected value is 0 under random classification, and a negative ARI value indicates classification that is worse than one would expect under random classification. Further details and discussion of the ARI are given by \cite{steinley04}.

\subsection{Alcoholic consumption data from the National Longitudinal Survey of Youth}
\subsubsection{The data}
The National Longitudinal Survey of Youth (NLSY) is a longitudinal study conducted by the United States Bureau of Labor Statistics with the goal of understanding the interaction between labor force participation, education, and health behaviors in children and adolescents. The sample for this study was a cohort of children who were between the ages of 12 and 17 when first interviewed in 1997. The data of interest were gathered each year between 1997 and 2011 and again in 2013 (15 total possible interviews). Each respondent provided a number between 0 and 98 that represents the number of alcoholic drinks they typically consume on a given day on which they are drinking. Because we are interested in modeling drinking behaviour over the life span, the data are shifted from representing year of interview to age. We follow individuals who were first interviewed at age 16 until they are 19; all individuals with missing inputs are excluded and no covariates are adopted. To this end, 1151 observations with four time points (i.e., at ages 16, 17, 18, and 19) are used for the following analyses. 

\subsubsection{Model selection}
We implement the Gaussian GMM via {\tt{Mplus}} Version 7.1 \citep{muthen98}. Our proposed GHD-GMM and GST-GMM are implemented in {\sf R} and run with $K=1,\ldots,10$ until the best model is obtained under each scenario. Table~\ref{nlsyresult} shows the results of fitting all of the models as aforementioned for a varying number of latent classes. The BIC values show that more than eight classes are needed with the conventional GMM, two are needed with constrained Models~I and~IV, and three are needed for all of the other models. The BIC values for the GST-GMM and GHD-GMM are always better than the BIC for the normal GMM. Notably, the BIC values for the GHD-GMM do not always improve on those for the GST-GMM. Among all fitted models, the three-cluster general GST-GMM under $\vecbeta_{yk}=\mathbf{0}$ (i.e., general Model III) is preferable according to the BIC. It is worth mentioning that, even though the skew-t distribution is a special case of the generalized hyperbolic distribution, the GST-GMM seems to be useful in addition to the GHD-GMM.
\begin{table}[!ht]
\caption{Results of fitting normal, GST, and GHD GMMs for consumption data from the National Longitudinal Survey of Youth.}
\label{nlsyresult}
	\centering
	%\resizebox{15.5cm}{!}{
	\small{
	%\footnotesize{
	%\scriptsize{
	\begin{tabular*}{1.0\textwidth}{@{\extracolsep{\fill}}lrrrrrr}
	\hline
	&\multicolumn{3}{c}{GMM--Normal (constrained)}  &\multicolumn{3}{c}{GMM--Normal (general)}\\
	\cline{2-4}\cline{5-7}
	Class & Log-likelihood & Free paras & BIC  & Log-likelihood & Free paras & BIC\\ 
\hline
	1 & --14983.42 & 9 & --30030.27 & --14983.42&  9     & --30030.27  \\  
	2 & --14623.41 & 12 & --29331.40 & --12671.88&  19     &  --25477.69\\ 
	3 & --14330.00 & 15 & --28765.72 & --12233.95&    29   & --24672.31  \\ 
	4 & --14182.66 & 18 & --28492.19 & --12119.21&     39  &  --24513.30 \\ 
	5 & --14076.42 & 21 & --28300.85 &  --12027.60   & 49      & --24400.58  \\ 
	6 & --14015.58 & 24 & --28200.32 &   --11950.97  & 59     &  --24317.78 \\ 
	7 & --13980.78 & 27 & --28151.86 &   --11906.09   &     69 & --24298.53  \\ 
	8 & --13937.17 & 30 & --28085.80 &   --11870.44  &     79 &--24297.69 \\ 
	9 & --13916.40 & 33 & --28065.38&                        &      &   \\ 
\hline
	&\multicolumn{3}{c}{GHD--GMM (Model I, constrained)} &\multicolumn{3} {c}{GHD--GMM (Model I, general)} \\
	\cline{1-4}\cline{5-7}
	Class & Log-likelihood & Free paras & BIC & Log-likelihood & Free paras & BIC \\ 
\hline
	1 & --12403.20 & 13 & --24898.19 & --12403.28 & 13 & --24898.19 \\ 
	2 & --12315.75 & 16 & --24744.27& --12119.53 & 27 & --24429.36 \\ 
	3 & --12315.50 & 19 & --24764.91 & --11958.92 & 41 & \textbf{--24206.82}\\ 
	4 &                     &      &                     & --11953.84 & 55 & --24295.33  \\ 
\hline
	&\multicolumn{3}{c}{GHD--GMM (Model II, constrained)} &\multicolumn{3}{c} {GHD--GMM (Model II, general)}  \\ 
	\cline{1-4}\cline{5-7}
	Class & Log-likelihood & Free paras & BIC & Log-likelihood & Free paras & BIC \\ 
\hline
	1 & --12399.68 & 15 & --24883.94 & --12399.68 & 15 & --24905.09 \\ 
	2 & --12312.27 & 18 & --24737.32 & --12166.47 & 31 & --24551.45 \\ 
	3 & --12288.26 & 21 & --24717.49 & --12002.12 & 47 & --24335.51 \\ 
	4 & --12287.98 & 24 & --24745.12 & --11956.47 & 63 & --24356.99 \\ 
\hline
	&\multicolumn{3}{c}{GST--GMM (Model III, constrained)} &\multicolumn{3} {c}{GST--GMM (Model III, general)}  \\ 
	\cline{1-4}\cline{5-7}
	Classes & Log-likelihood & Free paras & BIC & Log-likelihood & Free paras & BIC \\ 
\hline
	1 & --12421.85 & 12 & --24928.28 & --12421.92 & 12 & --24928.42 \\ 
	2 & --12352.31 & 15 & --24810.34 & --12151.6 & 25 & --24479.41 \\ 
	3 & --12340.61 & 18 & --24808.10 & --11966.84 & 38 &\textbf{ --24201.52} \\ 
	4 & --12348.28 & 21 & --24844.58 & --11925.67 & 51 & --24210.82 \\ 
\hline
	&\multicolumn{3}{c}{GST--GMM (Model IV, constrained)}&\multicolumn{3}{c} {GST--GMM (Model IV, general)}  \\ 
	\cline{1-4}\cline{5-7}
	Classes & Log-likelihood & Free paras & BIC & Log-likelihood & Free paras & BIC \\ 
\hline
	1 & --12418.18 & 14 & --24935.05 & --12418.19 & 14 & --24935.05 \\ 
	2 & --12348.01 & 17 & --24748.06 & --12118.12 & 29 & --24440.64 \\ 
	3 & --12347.48 & 20 & --24756.18 & --11990.60  & 44 & --24291.33 \\ 
	4 &                    &        &                   & --11938.08 & 59 & --24292.00 \\ 
	\hline
	\end{tabular*}}
\end{table}

\subsubsection{Interpretation of the best model}
The best-fitting model, the three-class Model III, breaks the data into three groups. %The classification table of the group membership with average number of drinks per drinking day is presented in Table~\ref{drinkperday}. 
From Table~\ref{AlcoholicParaEst}, it can be seen that Class 1 comprising 56\% of the population, begins with low-moderate drinking ($<1$ drink per drinking day), slightly increases during adolescence, and by age 19 the average drinks per drinking day is at about~1. These can be considered ``consistent low" drinkers. Although the intercept for this class is heavily positively skewed (intercept skewness $=2.59$), the slope is not (intercept skewness $=0.03$), which indicates that the individual slopes are nearly normally distributed around the class slope of 0.21. The second class, comprising 24\% of the population, are what will be called the ``decreasing" drinkers. This class has an intercept of around five drinks per drinking day (a drinking binge) and ends at about 3 drinks per drinking day (just below the amount considered a drinking binge).\footnote{The World Health Organization defines heavy episodic drinking (also called a drinking ``binge") as the consumption of 60 or more grams of alcohol on one occasion ({\tt www.who.int/gho/alcohol/consumption\_patterns/heavy\_episodic\_drinkers\_text/en/}), which is about four standard drinks ({\tt www.niaaa.nih.gov/alcohol-health/overview-alcohol-consumption/what-
standard-drink}).} The intercept is again positively skewed (intercept skewness $=2.90$) but the slope is negatively skewed (slope skewness $=-0.78$), suggesting that individuals in this class decrease their consumption quickly over the period of adolescence. The third class, comprising 20\% of the population, will be called the ``increasing moderate" drinkers. Their initial level of drinking is around 2.87 drinks per drinking day (less than a binge) and this increases during adolescence, ending at age 19 around 7 drinks per drinking day (far above a drinking binge). Both the slope and intercept are slightly positively skewed (intercept skewness $=0.48$, slope skewness $=0.41$; see Table~\ref{AlcoholicParaEst}).
\begin{table}[!ht]
\caption{Key parameter estimates for the best model (three-class Model III).}
\label{AlcoholicParaEst}
	\centering
	\small{
	%\footnotesize{
	\begin{tabular*}{1.0\textwidth}{@{\extracolsep{\fill}}lrrr}
		\hline
		&\multicolumn{3}{c}{Class}\\
\cline{2-4}
	&$k=1$&$k=2$&$k=3$\\
	\hline
	$n_k$&645&276&230\\
	$\sum_{i=1}^n\pi_{ik}$&0.56&0.24&0.20\\
	$\alpha_k$&$(0.32,0.21)^{'}$&$(4.98,-0.45)^{'}$&$(2.87,1.37)^{'}$\\
	$\beta_k$&$(2.58,0.03)^{'}$&$(2.90,-0.78)^{'}$&$(0.48,0.41)^{'}$\\
	$\nu_k$&6.83&2.97&2.62\\
	$\noisev_k$&$\left[\begin{array}{cc}0.22&-0.17\\-0.17&0.15\end{array}\right]$&$\left[\begin{array}{cc}0.10&-0.16\\-0.16&0.50\end{array}\right]$&$\left[\begin{array}{cc}0.14&-0.05\\-0.05&0.02\end{array}\right]$\\
	\hline
	\end{tabular*}}
\end{table}

These results suggest that, during adolescence, which is typically a time when alcohol consumption is initiated, individuals will have different reactions to the exposure to alcohol given their previous experience. Those individuals who are low drinkers will tend to continue to be low drinkers, those who have already consumed alcohol heavily will begin to taper back to safe levels (alluding to these individuals ``knowing their limits" when it comes to alcohol), and those who are only at moderate levels tend to increase to heavy drinking. This model may be useful because for indicating which 15-year-olds should be the target of interventions if the goal is to prevent heavy drinking in late adolescence. Although the high drinkers may appear to be the most likely to develop problems related to alcohol, they may ``grow out" of their alcohol consumption; most especially, the 15-year-olds that only drink at moderate levels should not be neglected. 

\citet{kerr00} find similar patterns using three different youth surveys. They measure the stability of alcohol consumption over four time periods. Time~1 cohort are separated into three classes: abstainers, moderate drinkers, and heavy drinkers. They find a high proportion of abstainers continue to abstain and very few drink more than once or drink heavily. Moderate drinkers also show considerable stability in these samples, with 70\% or more staying in the moderate category. Time~1 heavy drinkers are the least stable. In all three surveys, less than half of the heavy drinkers remain heavy drinkers.  
Our results also aligned better with those in \cite{warner07}, who found three groups of adolescent drinking initiation: a large group with no or low drinking (our ``consistent low" drinkers), a group that drank exclusively in adolescence and then decreased (analogous to our ``decreasing" drinkers), and a group that started low and increased (analogous to our ``increasing moderate" drinkers).

\subsubsection{Partition study}
Other models tend to find similar latent classes. For instance, the three-class Model~I (which has a very similar BIC to the three-class skew-t model) demonstrates a similar partition (Table~\ref{PartitionCompare}). This suggests that the same pattern endures regardless of the distributional assumptions. However, the cluster proportions differ slightly (58\%, 24\%, and 18\%, for the low, high/decreasing, and moderate/increasing classes, respectively), which seems to suggest that the GHD model classifies more individuals into the ``low'' class than the skew-t model. If the goal of the analysis is to identify groups to target for interventions for the prevention of alcoholism, the proportions found in the skew-t model might be preferred as they create population groups that are larger. Therefore, interventions targeting this group may have a greater impact on the population than those targeting a smaller group. 
\begin{table}[!ht]
\caption{Confusion matrix between the partition obtained by the 3-class Model I and the partition obtained by the 3-class Model III.}
\label{PartitionCompare}
	\centering
	\small{
	%\footnotesize{
	\begin{tabular*}{1.0\textwidth}{@{\extracolsep{\fill}}clccc}
		\hline
	&&\multicolumn{3}{c}{Model III, general}\\
	&&Consistent low&Decreasing &Increasing moderate\\
	\multirow{3}{*}{Model I, general}&Consistent low&645&0&23\\
	&Decreasing&0&276&0\\
	&Increasing moderate&0&0&207\\
		\hline
	\end{tabular*}}
\end{table}

\subsection{Simulation studies}
\begin{figure*}
        \centering
        \begin{subfigure}[b]{0.425\textwidth}
            \centering
            \includegraphics[width=\textwidth]{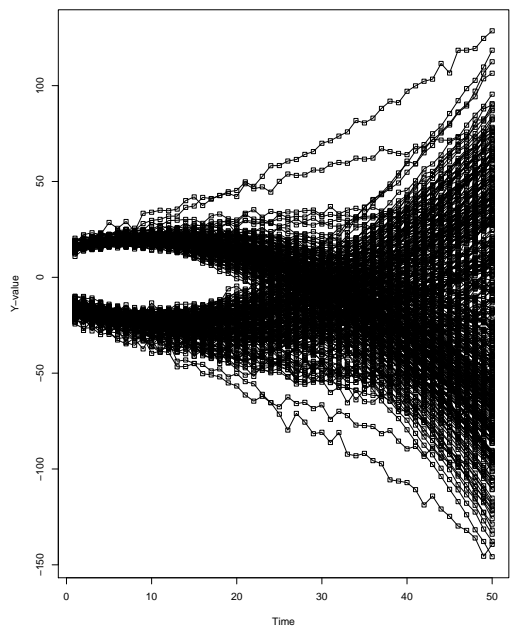}
            \caption[Network2]%
            {{\small Simulation 1}}    
            \label{fig:sim1}
        \end{subfigure}
        \quad
        \begin{subfigure}[b]{0.425\textwidth}  
            \centering 
            \includegraphics[width=\textwidth]{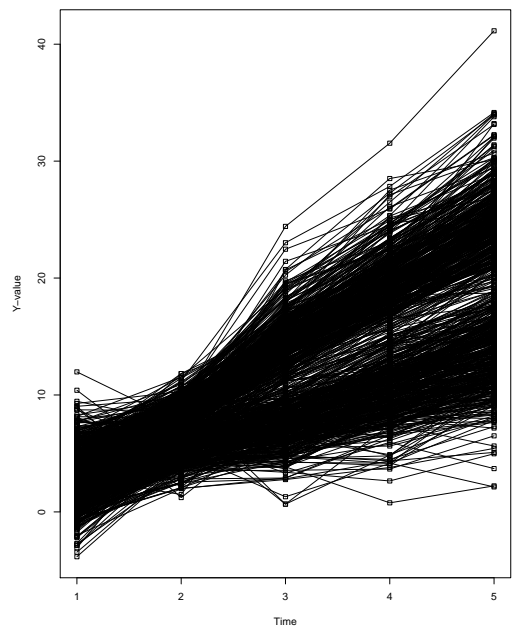}
            \caption[]%
            {{\small Simulation 2}}    
            \label{fig:sim2}
        \end{subfigure}
        \vskip\baselineskip
        \begin{subfigure}[b]{0.425\textwidth}   
            \centering 
            \includegraphics[width=\textwidth]{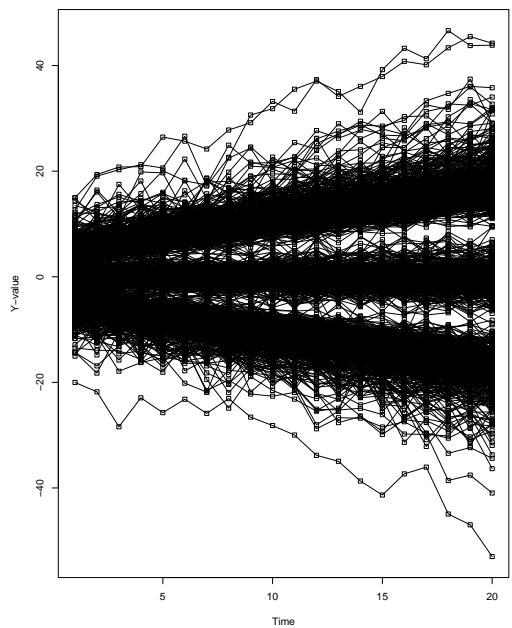}
            \caption[]%
            {{\small Simulation 3}}    
            \label{fig:sim3}
        \end{subfigure}
        \quad
        \begin{subfigure}[b]{0.425\textwidth}   
            \centering 
            \includegraphics[width=\textwidth]{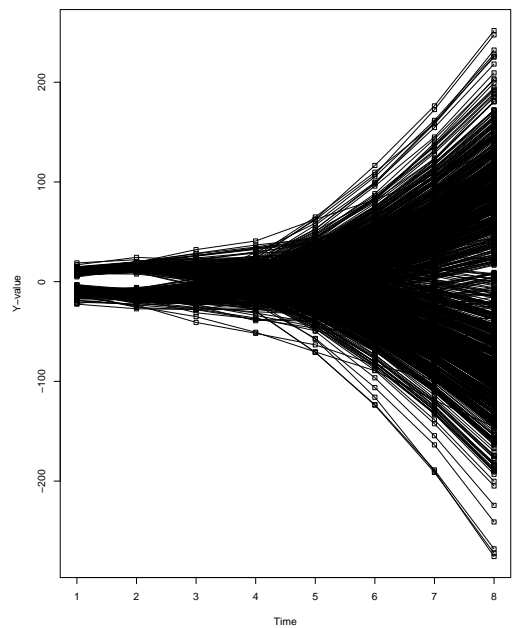}
            \caption[]%
            {{\small Simulation 4}}    
            \label{fig:sim4}
        \end{subfigure}
        \caption[ Simulation Examples]
        {\small Individual observation trajectories plots for the four simulation experiments.} 
        \label{fig:simutrajectory}
    \end{figure*}

In addition to real data application of our proposed model, we perform simulation studies with data generated in a number of scenarios: linear and quadratic GMMs with different distributions of the measurement errors and random effects, resulting in four distinct simulated data examples (see Table~\ref{SimDetail} for details). Individual trajectories for these four simulation experiments are plotted in Figure~\ref{fig:simutrajectory}. 
\begin{table}[!ht]
\caption{Key characteristics for Simulations~1--4.}
\label{SimDetail}
	\centering
	%\resizebox{15.5cm}{!}{
	\small{
	%\footnotesize{
	\begin{tabular*}{1.0\textwidth}{@{\extracolsep{\fill}}lrrrrrrr}
	\hline
Simulation&Model&$n$&$K$&$T$&$q$&$\sum_{i=1}^n\pi_{ik}$&Partition\\
\hline
1&Model I&400&2&50&3&$({1}/{2},{1}/{2})$&Overlapping\\	
2&Model II&800&2&5&2&$({1}/{2},{1}/{2})$&Separated\\
3&Model III&1500&3&20&2&$({1}/{3},{1}/{3},{1}/{3})$&Separated\\
4&Model IV&1000&2&8&3&$({1}/{2},{1}/{2})$&Overlapping\\	
\hline
\end{tabular*}}
\end{table}

We assess the performance of the family of non-elliptical GMMs in several different ways: Section~\ref{sec:sim1} illustrates the ability of our proposed family of models to recover underlying parameters when the number of classes and the model are correctly specified; then we present a comparison of the proposed models on BIC, ARI, and ERR from the clustering result for each simulation in Section~\ref{sec:sim2}; last but not least, 
we compare our proposed family of non-elliptical GMMs with Gaussian GMM developed by Muth{\'e}n and colleagues, which dominates the literature on GMMs in Section~\ref{sec:sim3} 
  
\subsubsection{Parameter recovery under the true model}\label{sec:sim1}
First, we evaluate the ability of our proposed model to recover underlying parameters when the number of classes and the model are correctly specified. To this end, 100 datasets are generated for each of the  four simulation experiments. True values and the means of the parameter estimates with their associated standard deviations are summarized in Tables~\ref{simu1result}--\ref{simu4result}. The results for each of the first three simulation experiments show that the means of all parameter estimates are close to the true values, with small standard deviations. The means of the last four elements of $\vecbeta_1$ and $\vecbeta_2$ are different from the true values due to the overlap between classes. %It is noteworthy to mention that, although the purpose of these simulation experiments was not to demonstrate clustering ability, each experiment consistently returned perfect clustering results.
\begin{table}[!ht]
\caption{Key model parameters as well as means and standard deviations of the associated parameter estimates from the 100 runs for the first simulation experiment.}
\label{simu1result}
	\centering
	\small{
	%\footnotesize{
	\begin{tabular*}{1.0\textwidth}{@{\extracolsep{\fill}}lrrr}
		\hline
	&True values&Means&Standard deviations\\
	\hline
	$\alpha_1$&$(15,8,-6)^{'}$&$(14.98,8.00,-6.00)^{'}$&$(0.10,0.09,0.15)^{'}$\\
	$\alpha_2$&$(-14,-10,6)^{'}$&$(-14.06,-9.96,6.02)^{'}$&$(0.24,0.20,0.16)^{'}$\\
	$\beta_1$&$(1,1,1)^{'}$&$(1.05,1.01,1)^{'}$&$(0.35,0.32,0.39)^{'}$\\
	$\beta_2$&$(-1,-1,-1)^{'}$&$(-0.89,-0.95,-0.95)^{'}$&$(0.24,0.25,0.27)^{'}$\\
	$\lambda_1$&-1&-0.52&0.70\\
	$\lambda_2$&2&1.05&1.28\\
	$\omega_1$&2&2.02&0.31\\
	$\omega_2$&3&2.92&0.51\\
	$\noisev_1$&$\left[ \begin{array}{ccc}1&0&0\\0&0.7&0\\0&0&2\end{array}\right]$&$\left[\begin{array}{ccc}0.99&0.00&0.00\\0.00&0.70&0.00\\0.00&0.00&1.99\end{array}\right]$&$\left[\begin{array}{ccc}0.33&0.07&0.12\\0.07&0.21&0.11\\0.12&0.11&0.62\end{array}\right]$\\
	$\noisev_2$&$\left[ \begin{array}{ccc}1.5&0&0\\0&0.8&0\\0&0&0.9\end{array}\right]$&$\left[\begin{array}{ccc}1.38&0.00&0.01\\0.00&0.74&-0.01\\0.01&-0.01&0.84\end{array}\right]$&$\left[\begin{array}{ccc}0.39&0.08&0.08\\0.08&0.21&0.08\\0.08&0.08&0.21\end{array}\right]$\\
	\hline
	\end{tabular*}}
\end{table}
\begin{table}[!ht]
\caption{Key model parameters as well as means and standard deviations of the associated parameter estimates from the 100 runs for the second simulation experiment.}
\label{simu2result}
	\centering
	\small{
	%\footnotesize{
	\begin{tabular*}{1.0\textwidth}{@{\extracolsep{\fill}}lrrr}
	\hline
	&True values&Means&Standard deviations\\
	\hline
	$\alpha_1$&$(4,5)^{'}$&$(4.00,5.00)^{'}$&$(0.11,0.08)^{'}$\\
	$\alpha_2$&$(2,3)^{'}$&$(1.99,2.98)^{'}$&$(0.18,0.11)^{'}$\\
	$\beta_1$&$(1,-1,1,1,1)^{'}$&$(0.94,-0.92,0.93,0.94,0.93)^{'}$&$(0.35,0.33,0.39,0.44,0.51)^{'}$\\
	$\beta_2$&$(-1,1,-1,-1,-1)^{'}$&$(-0.77,0.83,-0.72,-0.68,-0.67)^{'}$&$(0.27,0.36,0.33,0.43,0.55)^{}$\\
	$\lambda_1$&-1&-0.7&0.73\\
	$\lambda_2$&-2&-1.05&0.85\\
	$\omega_1$&2&2.02&0.32\\
	$\omega_2$&3&3.11&0.58\\
	$\noisev_1$&$\left[ \begin{array}{cc}1&0\\0&0.7\end{array}\right]$&$\left[\begin{array}{cc}0.92&0.00\\0.00&0.65\end{array}\right]$&$\left[\begin{array}{cc}0.34&0.08\\0.08&0.22\end{array}\right]$\\
	$\noisev_2$&$\left[ \begin{array}{cc}1.5&0\\0&0.8\end{array}\right]$&$\left[\begin{array}{cc}1.20&-0.01\\-0.01&0.63\end{array}\right]$&$\left[\begin{array}{cc}0.36&0.08\\0.08&0.18\end{array}\right]$\\
	\hline
	\end{tabular*}}
\end{table}
\begin{table}[!ht]
\caption{Key model parameters as well as means and standard deviations of the associated parameter estimates from the 100 runs for the third simulation experiment.}
\label{simu3result}
	\centering
	\small{
	%\footnotesize{
	\begin{tabular*}{1.0\textwidth}{@{\extracolsep{\fill}}lrrr}
		\hline
	&True values&Means&Standard deviations\\
	\hline
	$\alpha_1$&$(4,5)^{'}$&$(4.01,4.98)^{'}$&$(0.11,0.10)^{'}$\\
	$\alpha_2$&$(0,0)^{'}$&$(-0.01,0.01)^{'}$&$(0.07,0.08)^{'}$\\
	$\alpha_3$&$(-4,-5)^{'}$&$(-3.91,-4.9)^{'}$&$(0.81,1.01)^{'}$\\
	$\beta_1$&$(1,1)^{'}$&$(1.00,1.01)^{'}$&$(0.10,0.09)^{'}$\\
	$\beta_2$&$(0,0)^{'}$&$(-0.01,-0.02)^{'}$&$(0.17,0.19)^{'}$\\
	$\beta_3$&$(-1,-1)^{'}$&$(-0.99,-0.98)^{}$&$(0.24,0.22)^{'}$\\
	$\nu_1$&7&7.09&0.61\\
	$\nu_2$&5&4.97&0.41\\
	$\nu_3$&6&6.08&0.50\\
	$\noisev_1$&$\left[\begin{array}{cc}1&0\\0&0.7\end{array}\right]$&$\left[\begin{array}{cc}1.00&0.01\\0.01&0.68\end{array}\right]$&$\left[\begin{array}{cc}0.07&0.05\\0.05&0.07\end{array}\right]$\\
	$\noisev_2$&$\left[\begin{array}{cc}0.7&0\\0&0.6\end{array}\right]$&$\left[\begin{array}{cc}0.72&0.03\\0.03&0.64\end{array}\right]$&$\left[\begin{array}{cc}0.28&0.32\\0.32&0.40\end{array}\right]$\\
	$\noisev_3$&$\left[\begin{array}{cc}1.5&0\\0&0.8\end{array}\right]$&$\left[\begin{array}{cc}1.36&0.00\\0.00&0.76\end{array}\right]$&$\left[\begin{array}{cc}0.27&0.07\\0.07&0.08\end{array}\right]$\\
	\hline
	\end{tabular*}}
\end{table}
\begin{table}[!ht]
\caption{Key model parameters as well as means and standard deviations of the associated parameter estimates from the 100 runs for the fourth simulation experiment.}
\label{simu4result}
	\centering
	%\resizebox{15.5cm}{!}{
	\small{
	%\footnotesize{
	\begin{tabular*}{1.0\textwidth}{@{\extracolsep{\fill}}lrrr}
		\hline
	&True values&Means&Standard deviations\\
	\hline
	$\alpha_1$&$(8,7,-3)^{'}$&$(7.95,  7.01, -2.93)^{'}$&$(0.18,0.11,0.05)^{'}$\\
	$\alpha_2$&$(-8,-7,3)^{'}$&$(-7.98, -6.99,  2.93)^{'}$&$(0.21,0.12,0.05)^{'}$\\
	\multirow{2}{*}{$\beta_1$}&$(1,1,1,1,$&$(1.04,0.98,0.78,0.44,$&$(0.16,0.17,0.18,0.18,$\\
	                                                &$1,1,1,1)^{'}$&$-0.02,-0.62,-1.35,-2.21)^{'}$&$0.18,0.18,0.22,0.32)^{'}$\\
	\multirow{2}{*}{$\beta_2$}&$(-1,-1,-1,-1,$&$(-1.02,-0.96,-0.78,-0.46,$&$(0.17,0.15,0.16,0.16,$\\
	                                               &$-1,-1,-1,-1)^{'}$&$-0.01,0.56,1.26,2.09)^{'}$&$0.18,0.22,0.29,0.40)^{'}$\\
	$\nu_1$&7&7.43&0.80\\
	$\nu_2$&6&6.27&0.59\\
	$\noisev_1$&$\left[\begin{array}{ccc}1&0&0\\0&0.7&0\\0&0&0.8\end{array}\right]$&$\left[\begin{array}{ccc}0.99&0.00&0.01\\0.00&0.71&0.00\\0.01&0.00&0.80\end{array}\right]$&$\left[\begin{array}{ccc}0.11&0.07&0.05\\0.07&0.06&0.04\\0.05&0.04&0.06\end{array}\right]$\\
	$\noisev_2$&$\left[\begin{array}{ccc}1.5&0&0\\0&0.8&0\\0&0&0.9\end{array}\right]$&$\left[\begin{array}{ccc}1.49&0.01&0.01\\0.01&0.79&0.01\\0.01&0.01&0.90\end{array}\right]$&$\left[\begin{array}{ccc}0.20&0.10&0.08\\0.10&0.09&0.04\\0.08&0.04&0.07\end{array}\right]$\\
	\hline
	\end{tabular*}}
\end{table}

\subsubsection{Comparing Models I--IV}\label{sec:sim2}
Second, we compare Models I--IV. One hundred datasets are generated for the four simulation experiments above and analyzed using the GMMs developed herein. The means of the BIC, the ARI, and the ERR are summarized in Table~\ref{simucomparef}. For Simulations~1--3, the best models obtained are those with underlying true data structures, as expected. The BIC selects Model~2 with the correct number of components for Simulation~4; however, the estimated indices, i.e., $\hat\lambda_1=-2.93$ and $\hat\lambda_2=-2.70$, are very close to the parametrization under the skew-t distribution.
\begin{table}[!ht]
\caption{Comparison of results --- including average BIC, ARI, and ERR values --- for Models~I--IV, where bold face font is used to highlight the model with the best BIC for each simulation.}
\label{simucomparef}
	\begin{minipage}{\textwidth}
	\centering
	%\resizebox{15.5cm}{!}{
	\small{
	%\footnotesize{
	\begin{tabular*}{1.0\textwidth}{@{\extracolsep{\fill}}ccccccccc}
	\hline
%	Data structure&\multicolumn{8}{c}{Fitting models}\\
%	\cline{2-9}
	&\multicolumn{4}{c}{Model I} &\multicolumn{4}{c} {Model II} \\
	\cline{2-5}\cline{6-9}
	&Free paras&BIC & ARI & ERR &Free paras& BIC & ARI & ERR\\
\hline
	\multirow{2}{*}{Simulation 1}& \multirow{2}{*}{89}& \multirow{2}{*}{\textbf{--69895}} & \multirow{2}{*}{1.00} & \multirow{2}{*}{0.00}& \multirow{2}{*}{143}& \multirow{2}{*}{--70507} & \multirow{2}{*}{1.00} & \multirow{2}{*}{0.00}  \\ 
	&&&&&&\\
	\multirow{2}{*}{Simulation 2}& \multirow{2}{*}{29}& \multirow{2}{*}{--16018} & \multirow{2}{*}{0.92} & \multirow{2}{*}{0.02}& \multirow{2}{*}{35}& \multirow{2}{*}{\textbf{--15366}} & \multirow{2}{*}{0.97} & \multirow{2}{*}{0.01}  \\
	&&&&&&\\
	\multirow{2}{*}{Simulation 3}& \multirow{2}{*}{36}& \multirow{2}{*}{--108937} & \multirow{2}{*}{0.50} & \multirow{2}{*}{0.33}& \multirow{2}{*}{71}& \multirow{2}{*}{--106890} & \multirow{2}{*}{0.53} & \multirow{2}{*}{0.33}  \\ 
	&&&&&&\\
        \multirow{2}{*}{Simulation 4}& \multirow{2}{*}{69}& \multirow{2}{*}{--37133} & \multirow{2}{*}{1.00} & \multirow{2}{*}{0.00}& \multirow{2}      	{*}{103}& \multirow{2}{*}{\textbf{--36985}} & \multirow{2}{*}{1.00} & \multirow{2}{*}{0.00}  \\ 
	&&&&&&\\
\hline
	&\multicolumn{4}{c}{Model III} &\multicolumn{4}{c} {Model IV} \\
	\cline{2-5}\cline{6-9}
	&Free paras&BIC & ARI & ERR &Free paras& BIC & ARI & ERR \\
\hline
	\multirow{2}{*}{Simulation 1}& \multirow{2}{*}{87}& \multirow{2}{*}{--69945} & \multirow{2}{*}{1.00} & \multirow{2}{*}{0.00}& \multirow{2}{*}{139}& \multirow{2}{*}{--69946} & \multirow{2}{*}{1.00} & \multirow{2}{*}{0.00}  \\ 
	&&&&&&\\
	\multirow{2}{*}{Simulation 2}& \multirow{2}{*}{27}& \multirow{2}{*}{--15495} & \multirow{2}{*}{0.87} & \multirow{2}{*}{0.04}& \multirow{2}{*}{33}& \multirow{2}{*}{--16110} & \multirow{2}{*}{0.86} & \multirow{2}{*}{0.04}  \\ 
	&&&&&&\\

	\multirow{2}{*}{Simulation 3}& \multirow{2}{*}{33} & \multirow{2}{*}{--101669} & \multirow{2}{*}{1.00} & \multirow{2}{*}{0.00} &  \multirow{2}{*}{68}& \multirow{2}{*}{\textbf{--101587}} & \multirow{2}{*}{1.00} & \multirow{2}{*}{0.00} \\
	&&&&&&\\
	\multirow{2}{*}{Simulation 4}& \multirow{2}{*}{67} & \multirow{2}{*}{--37271} & \multirow{2}{*}{1.00} & \multirow{2}{*}{0.00} & \multirow{2}{*}{101} & \multirow{2}{*}{--37265} & \multirow{2}{*}{1.00} & \multirow{2}{*}{0.00} \\ 
	&&&&&&\\
	\hline
	\end{tabular*}}
	\end{minipage}
\end{table}

\subsubsection{Comparison with Gaussian GMMs}\label{sec:sim3}
Finally, we compare our proposed family of models with the Gaussian GMMs (via {\tt{Mplus}}). First, 100 datasets are generated, as described before, from Simulation~1 where the distributions of random effects are not normal. Table~\ref{percentBIC} summarizes the percentage of the replications favoured by the BIC when analyzing those 100 generated datasets for 1--6 latent classes (note that 6 latent classes were never selected, see Table~\ref{percentBIC}) via Models~I and~III as well as Gaussian GMMs. We then generate 100 datasets from Simulation~2 where the distribution of measurement errors are not normal (Table~\ref{percentBIC4}). It is not surprising that the Gaussian GMMs overestimate the number of classes in both cases because the normality assumptions are violated.  Moreover, when the normality assumption of the random effects is violated (Simulation 1), Gaussian GMMs tend to point to even more classes. It is noteworthy to mention that the best models, based on the BIC, are consistently Models~I and~II, respectively. 
\begin{table}[!ht]
\caption{Percent preferred by the BIC when analyzing the Simulation 1 with Model I, Model III, and GMM along with number of classes.}
%\vspace{0.1in}
\label{percentBIC}
	\centering
	\begin{tabular*}{1.0\textwidth}{@{\extracolsep{\fill}}lccccc}
	\hline   
	&\multicolumn{5}{c}{Number of classes}\\
	\cline{2-6}
	&1&2&3&4&5\\
	\hline
	Model I&0&100&0&0&0\\
	Model III&0&100&0&0&0\\
	GMM&0&0&0&67&33\\
	\hline
	\end{tabular*}
\end{table}
%\begin{table}[!ht]
%\caption{Results of fitting the Gaussian, Model I and III to Simulation 1.}
%\vspace{0.1in}
%\label{simutypical}
%	\centering
%	\begin{tabular*}{1.0\textwidth}{@{\extracolsep{\fill}}lccccrr}
%	\hline   
%	Model&$K$&Free paras&Log-likelihood&BIC&ARI&ERR\\
%	& ($K$)&&&&\\
%	\hline
%	&1&64&$-35261$&$-70906$&0&0.50\\
%	Model I&2&129&$-34283$&$\textbf{-69339}$&1&0\\
%	&3&194&$-34231$&$-69624$&0.86&0.08\\
%	\hline
%	&1&63&$-35295$&$-70967$&0&0.5\\
%	Model III&2&127&$-34292$&$-69346$&1&0\\
%	&3&191&$-34240$&$-69624$&0.86&0.09\\
%	\hline
%	&1&59&$-38474$&$-77303$&0&0.50\\
%	&2&119&$-35524$&$-71762$&0.77&0.06\\
%	GMM&3&179&$-34872$&$-70817$&0.51&0.29\\
%	&4&239&$-34544$&$-70521$&0.52&0.39\\
%	&5&299&$-34356$&$-70505$&0.46&0.45\\
%	&6&359&$-34252$&$-70656$&0.44&0.46\\
%	\hline
%	\end{tabular*}
%\end{table}

\begin{table}[!ht]
\caption{Percent preferred by the BIC when analyzing the Simulation 4 with Model II, Model IV, and GMM along with number of classes.}
%\vspace{0.1in}
\label{percentBIC4}
	\centering
	\begin{tabular*}{1.0\textwidth}{@{\extracolsep{\fill}}lccccc}
	\hline   
	&\multicolumn{5}{c}{Number of classes}\\
	\cline{2-6}
	&1&2&3&4&5\\
	\hline
	Model II&0&100&0&0&0\\
	Model IV&0&100&0&0&0\\
	GMM&0&0&24&57&19\\
	\hline
	\end{tabular*}
\end{table}

%%%%%%Discussion and Conclusion
\section{Discussion}\label{sec:discussion}
We have introduced novel GHD-GMM and GST-GMM models, which are extensions of the GMMs introduced by \citet{verbeke96} to the generalized hyperbolic and skew-t distributions, respectively, to facilitate heavier tails or asymmetry. Updates are derived for parameter estimation within the EM algorithm framework, which is made feasible by the fact that the generalized hyperbolic distribution can be represented as a normal mean-variance mixture, where the weight follows a GIG distribution. In our GMM extensions, four models were considered (GHD-GMM and GST-GMM under $\vecbeta_{yk} = \mathbf{0}$, GHD-GMM and GST-GMM under $\vecbeta_{\eta k} = \mathbf{0}$) and their performance was compared using simulated and real data. In terms of interpretation, GHD-GMM under $\vecbeta_{\eta k} = \mathbf{0}$ is preferable to GHD-GMM under $\vecbeta_{yk} = \mathbf{0}$ because the skewness parameters are in the data space and so the interpretation of the skewness parameters is clear. However, in terms of model complexity, GHD-GMM under $\vecbeta_{yk} = \mathbf{0}$ is preferable to GHD-GMM under $\vecbeta_{\eta k} = \mathbf{0}$ because the former model has $K(T-q)$ fewer parameters than the latter. 

We believe that this kind of mixture modeling approach for longitudinal data is important in many biostatistical and psychological applications, allowing accurate inference of model parameters and class membership probabilities while adjusting for heterogeneity, heavy tails, and skewness in the data. The proposed GHD-GMM and GST-GMM models have several advantages over Gaussian GMMs. The proposed GHD-GMM, which includes the multivariate skew-t, variance-gamma distribution, multivariate normal inverse-Gaussian distributions, etc., as special or limiting cases, provides flexibility to handle a broader range of multivariate longitudinal data --- the same is true, albeit to a lesser extent, for the proposed GST-GMM. In the presence of heterogeneity, heavy tails, and skewness in longitudinal data, the proposed models can fit the data considerably better than Gaussian mixtures thereby reducing the risk of extracting latent classes that are merely due to non-normality of the outcomes. When data are normal, the proposed GHD-GMM can be used to check the reproducibility of a Gaussian GMM solution due to the flexibility of the generalized hyperbolic distribution --- again, the same is true for the GST-GMM. However, when there  exist  outliers, while the concentration parameter mitigates the outliers, it has been shown that a contaminated approach can be more effective in handling the impact of the outliers \citep[see][]{punzo16}. Therefore, developing a contaminated version of the family of non-elliptical GMMs will be considered in future work.

The models proposed herein can also be further developed in various ways. For example, for the first level of the GMM, only $q$th order polynomial equations are considered but kernel regressions or non-linear regressions could be incorporated into the model. Bayesian mixture modeling may offer researchers an alternative way to handle clustering of longitudinal data due to the popularity and advances in Markov chain Monte Carlo techniques. Finally, it is also worthwhile to other flexible parametric distributions of measurement errors and random effects, such as the coalesced generalized hyperbolic distribution and the multiple scaled generalized hyperbolic distribution \citep{tortora16} and then hidden truncation hyperbolic distribution \citep{murray17b}. 

%\newpage
{\small%\bibliographystyle{chicago}
%\bibliography{gmmrefers}

}

%\newpage
\appendix
\section{Distribution of $\veceta_i~|~\vecy_i, \vecx_i, w_{ik},c_{ik}=1$} 
%\section{Appendix A: Conditional distribution of $\veceta$ given $\vecy, \vecx, w$ and $c$} 
Herein, we give the detailed derivation of the conditional distribution of $\veceta_i$ given $\vecy_i, \vecx_i, w_{ik}$, and $c_{ik}=1$, which facilitates computation of the conditional expectations in the E-step of the EM algorithm. It also serves as a way to estimate the growth factor scores. The joint distribution of $\veceta_i$ and $\vecY_i$ given $\vecx_i, w_{ik}$, and $c_{ik}=1$ is given by
\begin{equation*}
\begin{pmatrix} \veceta_i \\ \vecY_i \end{pmatrix} \bigg|~\vecx_i, w_{ik}, c_{ik}=1\sim \mathcal{N}\biggr(\begin{pmatrix} \vecalpha_k + \mGamma_k\vecx_i + w_{ik}\vecbeta_{\eta k} \\ \mLambda_y(\vecalpha_k + \mGamma_k\vecx_i + w_{ik}\vecbeta_{\eta k}) \end{pmatrix}, \begin{pmatrix} w_{ik}\noisev_k&w_{ik}\noisev_k\mLambda_y^{'} \\w_{ik}\mLambda_y\noisev_k&w_{ik}(\mLambda_y\noisev_k\mLambda_y^{'}+\vecTheta_k) \end{pmatrix} \biggr).
\end{equation*}
According to the properties of the conditional distribution for multivariate normal variables, the conditional distribution of $\veceta_i$ given $\vecy_i, \vecx_i, w_{ik}, c_{ik}=1$ is also a multivariate normal distribution with 
\begin{align*}
\E(\veceta_i \mid \vecy_i, \vecx_i, w_{ik}, c_{ik}=1 ) &= \vecalpha_k + \mGamma_k\vecx_i + w_{ik}\vecbeta_{\eta k} \\
&\quad+ \noisev_k\mLambda_y^{'}\inv{(\mLambda_y\noisev_k\mLambda_y^{'}+\vecTheta_k)}(\vecy_i - \mLambda_y(\vecalpha_k + \mGamma_k\vecx_i + w_{ik}\vecbeta_{\eta k})),\\
\mathbb{V}\text{ar}(\veceta_i \mid \vecy_i, \vecx_i, w_{ik}, c_{ik}=1 ) &= w_{ik}\noisev_k - w_{ik}\noisev_k\mLambda_y^{'} \inv{(\mLambda_y\noisev_k\mLambda_y^{'}+\vecTheta_k)} \mLambda_y\noisev_k.
\end{align*}
According to the Woodbury matrix identity \citep{woodbury50}, the covariance matrix for the latent variable $\veceta_i$ can be simplified to
\begin{equation*}
\mathbb{V}\text{ar}(\veceta_i \mid \vecy_i, \vecx_i, w_{ik}, c_{ik}=1 ) = w_{ik}\inv{(\inv{\noisev_k}+\mLambda_y^{'}\inv{\vecTheta_k}\mLambda_y)}.
\end{equation*}
Next, let us simplify the expectation of the latent variable $\veceta$: 
\begin{align*}
\E&(\veceta_i \mid \vecy_i, \vecx_i, w_{ik}, c_{ik}=1 )\\
&=(\ident_T -  \noisev_k\mLambda_y^{'}\inv{(\mLambda_y\noisev_k\mLambda_y^{'}+\vecTheta_k)}\mLambda_y)( \vecalpha_k + \mGamma_k\vecx_i + w_{ik}\vecbeta_{\eta k})
+ \noisev_k\mLambda_y^{'}\inv{(\mLambda_y\noisev_k\mLambda_y^{'}+\vecTheta_k)}\vecy_i,\\
&=(\noisev_k -  \noisev_k\mLambda_y^{'}\inv{(\mLambda_y\noisev_k\mLambda_y^{'}+\vecTheta_k)}\mLambda_y\noisev_k)\inv{\noisev_k}( \vecalpha_k + \mGamma_k\vecx_i + w_{ik}\vecbeta_{\eta k}) \\
&\quad+ \noisev_k\mLambda_y^{'}(\inv{\vecTheta_k}-\inv{\vecTheta_k}\mLambda_y\inv{(\inv{\noisev_k}+\mLambda_y^{'}\inv{\vecTheta_k}\mLambda_y)}\mLambda_y^{'}\inv{\vecTheta_k})\vecy_i,\\
&=\inv{(\inv{\noisev_k}+\mLambda_y^{'}\inv{\vecTheta_k}\mLambda_y)}\inv{\noisev_k}( \vecalpha_k + \mGamma_k\vecx_i + w_{ik}\vecbeta_{\eta k}) \\
&\quad+ (\noisev_k(\inv{\noisev_k}+\mLambda_y^{'}\inv{\vecTheta_k}\mLambda_y)- \noisev_k\mLambda_y^{'}\inv{\vecTheta_k}\mLambda_y)\inv{(\inv{\noisev_k}+\mLambda_y^{'}\inv{\vecTheta_k}\mLambda_y)}\mLambda_y^{'}\inv{\vecTheta_k}\vecy_i,\\
&=\inv{(\inv{\noisev_k}+\mLambda_y^{'}\inv{\vecTheta_k}\mLambda_y)}\inv{\noisev_k}( \vecalpha_k + \mGamma_k\vecx_i + w_{ik}\vecbeta_{\eta k}) + \inv{(\inv{\noisev_k}+\mLambda_y^{'}\inv{\vecTheta_k}\mLambda_y)}\mLambda_y^{'}\inv{\vecTheta_k}\vecy_i,\\
&=\inv{(\inv{\noisev_k}+\mLambda_y^{'}\inv{\vecTheta_k}\mLambda_y)}(\inv{\noisev_k}( \vecalpha_k + \mGamma_k\vecx_i + w_{ik}\vecbeta_{\eta k}) +\mLambda_y^{'}\inv{\vecTheta_k}\vecy_i).
\end{align*}
Finally, we obtain the conditional distribution
\begin{equation*}
\veceta_i \mid \vecy_i, \vecx_i, w_{ik}, c_{ik}=1\sim \mathcal{N}(\vecV_k(\inv{\noisev_k}( \vecalpha_k + \mGamma_k\vecx_i + w_{ik}\vecbeta_{\eta k}) +\mLambda_y^{'}\inv{\vecTheta_k}\vecy_i), w_{ik}\vecV_k),
\end{equation*}
where $\vecV_k =\inv{(\inv{\noisev_k}+\mLambda_y^{'}\inv{\vecTheta_k}\mLambda_y)}$.

\section{Detailed Parameter Estimation}
\subsection{EM algorithm for Model I} %(i.e., GHD-GMM under $\vecbeta_{yk}=\mathbf{0}$)}
\label{sec:modelipara}
For our GHD-GMM under $\vecbeta_{yk}=\mathbf{0}$, the observed log-likelihood can be expressed as follows:
\begin{equation}
\tlog \mathcal{L} = \lsum \log p(\vecy_i\mid\vecx_i),
\end{equation}
where 
\begin{equation}
p(\vecy_i\mid\vecx_i) = \ksum \pi_{ik} f_{\text{GHD}_T}(\vecy_i; \lambda_k,\omega_k,\vecmu_k,\matsig_k,\mLambda_y\vecbeta_{\eta k}).
\end{equation}
Now, $\vecY_i \mid\vecx_i,w_{ik},c_{ik}=1 \sim \mathcal{N}(\vecmu_k+ w_{ik}\mLambda_y \vecbeta_{\eta k}, w_{ik}\matsig_k)$ independently for $i = 1, \ldots,n$, $W_{ik}\mid c_{ik}=1\sim \mathcal{I} (\omega_k, 1, \tilde{\lambda}_k)$; therefore, from Bayes's theorem, $W_{ik}\mid\vecy_i,\vecx_i,c_{ik}=1\sim \text{GIG}(\psi_k,\chi_{ik},\tilde{\lambda}_{k})$ with $\psi_k = \omega_k +\vecbeta_{\eta k}^{'}\mLambda_y^{'}\inv\matsig_k\mLambda_y\vecbeta_{\eta k}$, $\chi_{ik} = \omega_k +\delta (\vecy_i,\vecmu_k\mid\matsig_k)$, and $\tilde{\lambda}_{k} = \lambda_k-T/2$. It follows that %Also, from the above equations, the conditional distribution of $\veceta_i$ given $\vecy_i$, $\vecx_i$, $w_{ik}$, and $c_{ik}=1$ is then
\begin{equation}
\label{eqn:eta2}
\veceta_i\mid\vecy_i,\vecx_i, w_{ik}, c_{ik}=1 \sim \mathcal{N}(\vecV_k(\inv{\noisev_k}(\vecalpha_k+\mGamma_{k}\vecx_i+w_{ik}\vecbeta_{\eta k})+\mLambda_y^{'}\inv{\vecTheta_k}\vecy_i),w_{ik}\vecV_k ),
\end{equation}
where $\vecV_k=\inv{(\inv{\noisev_k}+\mLambda_y^{'}\inv{\vecTheta_k}\mLambda_y)}$. The result in \eqref{eqn:eta2} is used to estimate the latent growth factors $\veceta_i$ and a detailed proof thereof is given in Appendix~A.
Therefore, the complete-data likelihood is given by
\begin{equation*}
\mathcal{L}_{\text{c}}(\varthet) = \lprod\kprod[\pi_{ik}\phi(\vecy_i\mid\mLambda_y\veceta_i,w_{ik}\vecTheta_k)\phi(\veceta_i\mid\vecalpha_k+\mGamma_{k}\vecx_i+w_{ik}\vecbeta_{\eta k},w_{ik}\noisev_k)h(w_{ik}\mid\omega_k,\lambda_k)]^{c_{ik}},
\end{equation*}
with the same notation used previously, %and where $\phi(\vecy_{i}\mid\mLambda_y\veceta_i,w_{ik}\vecTheta_k)$ is a multivariate Gaussian distribution with mean $\mLambda_y\veceta_i$ and covariance matrix $w_{ik}\vecTheta_k$, where $\phi(\veceta_i\mid\vecalpha_k+\mGamma_{k}\vecx_i,w_{ik}\noisev_k)$ is a multivariate Gaussian distribution with mean $\vecalpha_k+\mGamma_{k}\vecx_i$ and covariance matrix $w_{ik}\noisev_k$, and 
where $h(w_{ik}\mid\omega_k,\lambda_k)$ is the density of the GIG distribution in \eqref{eqn:gig} with $\eta = 1$. 
After some algebra, the complete-data log-likelihood is
\begin{equation}
\mathcal{L}_c (\varthet \mid \vecy, \vecx) =\mathcal{L}_{1c}(\mathbf{\pi}) + \mathcal{L}_{2c}(\vecTheta_k)+\mathcal{L}_{2c}(\vecalpha_k,\vecbeta_{\eta k},\noisev_k,\mGamma_{k})+\mathcal{L}_{4c}(\veclambda,\vecomega),
\end{equation}
where $\veclambda = (\lambda_1,\ldots,\lambda_K)$ and $\vecomega = (\omega_1,\ldots,\omega_K)$, and
\begin{align*}
\mathcal{L}_{1c}&=\lsum\ksum c_{ik}\tlog\pi_{ik},\\
\mathcal{L}_{2c}&=\lsum\ksum c_{ik}\left\{\frac{1}{2}\tlog|\vecTheta_k^{-1}| -  \frac{1}{2w_{ik}}\vecy_i^{'}{\vecTheta_k^{-1}}\vecy_i+\frac{1}{w_{ik}}\vecy_i^{'}{\vecTheta_k^{-1}}\mLambda_y\veceta_i-\frac{1}{2w_{ik}}\veceta^{'}\mLambda_y^{'}{\vecTheta_k^{-1}}\mLambda_y\veceta\right\}+B_1,\\  
\mathcal{L}_{3c}&=\lsum\ksum c_{ik}\left\{\frac{1}{2}\tlog|\noisev_k^{-1}| -  \frac{1}{2w_{ik}}\veceta_i^{'}{\noisev_k^{-1}}\veceta_i+\frac{1}{w_{ik}}(\vecalpha_k+\mGamma_{k}\vecx_i)^{'}{\noisev_k^{-1}}\veceta_i+\vecbeta_{\eta k}^{'}\noisev_k^{-1}\veceta_i\right.\\
&\left.\quad\quad\quad\quad-\frac{1}{2w_{ik}}(\vecalpha_k+\mGamma_{k}\vecx_i)^{'}{\noisev_k^{-1}}(\vecalpha_k+\mGamma_{k}\vecx_i)-(\vecalpha_k+\mGamma_{k}\vecx_i)^{'}\noisev_k^{-1}\vecbeta_{\eta k}-\frac{w_{ik}}{2}\vecbeta_{\eta k}^{'}\noisev_k^{-1}\vecbeta_{\eta k}\right\}+B_2,\\                     
\mathcal{L}_{4c}&=\lsum\ksum c_{ik}\left\{(\lambda_k-1)\log w_{ik} - \tlog K_{\lambda_k}(\omega_k) - \frac{\omega_k}{2}\left(w_{ik}+\frac{1}{w_{ik}}\right)\right\},
\end{align*}
where $B_1$ and $B_2$ are constants with respect to model parameters. 

In the E-step, we compute the conditional exception of $\mathcal{L}_c(\varthet \mid \vecy, \vecx)$ given in (32), denoted $\mathcal{Q}$.
%
%In the E-step, we compute the conditional expectation of $\mathcal{L}_c $ given the observed outcomes $\vecy_i$, the observed covariates $\vecx_i$, and the current parameter updates $\varthet^{\text{cur}}$:
%\begin{equation}
%\mathcal{Q} = %\mathbb{E} \left[ \mathcal{L}_c (\varthet ),\mathcal{L}( \varthet^{cur})\right] = 
%\mathbb{E} \left[ \mathcal{L}_c (\varthet \mid \vecy, \vecx); \varthet^{\text{cur}}  \right].
%\end{equation}
First, let $p_{ik}$ denote the probability that the $i$th observation belongs to the $k$th component of the mixture, and is updated by
\begin{equation*}
p_{ik}\colonequals\mathbb{E}\left[C_{ik} \mid \vecy_i,\vecx_i\right]=\frac{\pi_{ik} f_{\text{GHD}_T}(\vecy_i;\lambda_k,\omega_k,\vecmu_k,\matsig_k,\mLambda_y\vecbeta_{\eta k})}{\sum_{l=1}^{K} \pi_{il} f_{\text{GHD}_T}(\vecy_i; \lambda_l,\omega_l,\vecmu_l,\matsig_l,\mLambda_y\vecbeta_{\eta l})}.
\end{equation*}
The following expectations are required:
\begin{align*}
E_{1ik}&\colonequals\Ew=\sqrt{\frac{\chi_{ik}}{\psi_k}}\frac{K_{\tilde{\lambda}_k+1}(\sqrt{\psi_{k}\chi_{ik}})}{K_{\tilde{\lambda}_k}(\sqrt{\psi_{k}\chi_{ik}})},\\
\vspace{0.1in}
E_{2ik}&\colonequals\Ewinv=\sqrt{\frac{\psi_k}{\chi_{ik}}}\frac{K_{\tilde{\lambda}_k+1}(\sqrt{\psi_{k}\chi_{ik}})}{K_{\tilde{\lambda}_k}(\sqrt{\psi_{k}\chi_{ik}})}-\frac{2\tilde{\lambda}_k}{\chi_{ik}},\\
\vspace{0.1in}
E_{3ik}&\colonequals\Elogw=\log\left(\sqrt{\frac{\chi_{ik}}{\psi_k}}\right)+\frac{1}{K_{\tilde{\lambda}_k}(\sqrt{\psi_{k}\chi_{ik}})}\frac{\partial}{\partial \tilde{\lambda}_k}K_{\tilde{\lambda}_k}(\sqrt{\psi_{k}\chi_{ik}}),\\
\vspace{0.1in}
E_{4ik}&\colonequals\Eeta=\vecV_k(\noisev_k^{-1}(\vecalpha_k+\mGamma_{k}\vecx_i+E_{1ik}\vecbeta_{\eta k})+\mLambda_y^{'}\vecTheta_k^{-1}\vecy_i),\\
\vspace{0.1in}
E_{5ik}&\colonequals\Eetawinv=E_{2ik}\vecV_k(\noisev_k^{-1}(\vecalpha_k+\mGamma_{k}\vecx_i)+\mLambda_y^{'}\vecTheta_k^{-1}\vecy_i)+\vecV_k\noisev_k^{-1}\vecbeta_{\eta k},\\
\vspace{0.1in}
E_{6ik}&\colonequals\Eetaetawinv=\vecV_k+\vecV_k(\noisev_k^{-1}(\vecalpha_k+\mGamma_{k}\vecx_i)+\mLambda_y^{'}\vecTheta_k^{-1}\vecy_i)\vecbeta_{\eta k}\noisev_k^{-1}\vecV_k\\
&+E_{2ik}\vecV_k(\noisev_k^{-1}(\vecalpha_k+\mGamma_{k}\vecx_i)+\mLambda_y^{'}\vecTheta_k^{-1}\vecy_i)(\noisev_k^{-1}(\vecalpha_k+\mGamma_{k}\vecx_i)+\mLambda_y^{'}\vecTheta_k^{-1}\vecy_i)^{'}\vecV_k\\
&+\vecV_k\noisev_k^{-1}\vecbeta_{\eta k}^{'}(\noisev_k^{-1}(\vecalpha_k+\mGamma_{k}\vecx_i)+\mLambda_y^{'}\vecTheta_k^{-1}\vecy_i)^{'}\vecV_k+E_{1ik}\vecV_k\noisev_k^{-1}\vecbeta_{\eta k}\vecbeta_{\eta k}^{'}\noisev_k^{-1}\vecV_k,
\end{align*}
where $\psi_k$, $\chi_{ik}$, and $\tilde{\lambda}_k$ are as previously defined. These attractive closed forms for $E_{1ik}$, $E_{2ik}$, and $E_{3ik}$ exist because $W_{ik}\mid\vecy_i,\vecx_i,c_{ik}=1\sim \text{GIG}(\psi_k,\chi_{ik},\tilde{\lambda}_{k})$, and so we can use the formulae in \eqref{eqn:expectedvalue}. The exisitance of these attractive closed forms for $E_{4ik}$, $E_{5ik}$, and $E_{6ik}$ is due to the conditional Gaussian distribution of $\veceta$ as in \eqref{eqn:eta2}.

In the M-step, we maximize $\mathcal{Q}$ with respect to the model parameters to get the updates. In particular, we have to maximize 
\begin{equation}
\lsum\ksum p_{ik}\tlog\pi_{ik},
\end{equation}
with  respect to $\pi_{ik}$ for $k=1,\ldots, G$ and we obtain
\begin{equation*}
\hat{\pi}_{ik}=\frac{p_{ik}}{\sum_{k=1}^K p_{ik}}.
\end{equation*}
The updates for $\omega_k$ and $\lambda_k$ are computed by maximizing the following function
\begin{equation}
q_k(\omega_k,\lambda_k) = -\tlog K_{\lambda_k}(\omega_k) + (\lambda_k-1)\bar{d}_k-\frac{\omega_k}{2}(\bar{a}_k+\bar{b}_k),
\end{equation}
where $n_k=\lsum p_{ik}$, $\bar{a}_k=({1}/{n_k})\lsum p_{ik}E_{1ik}$, $\bar{b}_k=({1}/{n_k})\lsum p_{ik}E_{2ik}$, and $\bar{d}_k=({1}/{n_k})\lsum p_{ik}E_{3ik}$. The associated updates are
\begin{align} 
\hat{\lambda}_k&=\bar{c}_k\hat{\lambda}_k^\text{prev}\left[\frac{\partial}{\partial t}\tlog K_t(\hat{\omega}_k^{\text{prev}})\mid_{t=\hat{\lambda}_k^{\text{prev}}}\right]^{-1},\nonumber\\
\hat{\omega}_k&=\hat{\omega}_k^\text{prev}-\left[\frac{\partial}{\partial t}q_k(t,\hat{\lambda}_k)\Big|_{t=\hat{\omega}_k^{\text{prev}}}\right]\left[\frac{\partial^2}{\partial t^2}q_k(t,\hat{\lambda}_k)\Big|_{t=\hat{\omega}_k^{\text{prev}}}\right]^{-1},\nonumber
\end{align}
where the superscript ``prev'' means the previous estimate --- refer to \citet{browne15} for further details.
Finally, we get the updates of the other parameters in the model:
\begin{align*}
\hat{\vecTheta}_k&= \diag\biggr\{\frac{\lsum p_{ik}(E_{2ik}\vecy_i\vecy_i^{'} - \vecy_iE_{5ik}^{'}\mLambda_y^{'}-\mLambda_yE_{5ik}\vecy_i^{'}+\mLambda_yE_{6ik}\mLambda_y^{'})}{n_k}\biggr\},\\
\hat{\mGamma}_{k}&=\lsum p_{ik}\left(E_{5ik}\vecx_i^{'}-E_{2ik}\hat{\vecalpha}_k\vecx_i^{'}-\hat{\vecbeta}_k\vecx_i^{'}\right)\left(\lsum\ksum p_{ik} E_{2ik} \vecx_i\vecx_i^{'}\right)^{-1},\\
\hat{\vecalpha}_k&=\frac{\bar{a}_k\lsum p_{ik}(E_{5ik}-E_{2ik}\hat{\mGamma}_{k}\vecx_i)-\lsum p_{ik}E_{4ik} + \lsum p_{ik}\hat{\mGamma}_{k}\vecx_i}{n_k(\bar{a}_k\bar{b}_k-1)},\\
\hat{\vecbeta}_{\eta k}&=\frac{\bar{b}_k\lsum p_{ik}(E_{4ik}-\hat{\mGamma}_{k}\vecx_i)-\lsum p_{ik}E_{5ik} + \lsum p_{ik}E_{2ik}\hat{\mGamma}_{k}\vecx_i}{n_k(\bar{b}_k\bar{b}_k-1)},\\
\hat{\noisev}_k&=\frac{1}{n_k} \lsum p_{ik}\Big[E_{6ik}-\hat{\vecbeta}_{\eta k}E_{4ik}^{'}-E_{5ik}(\hat{\vecalpha}_k+\hat{\mGamma}_{k}\vecx_i)^{'}-E_{4ik}\hat{\vecbeta}_{\eta k}^{'}-(\hat{\vecalpha}_k+\hat{\mGamma}_{k}\vecx_i)E_{5ik}^{'}\\
&+E_{2ik}(\hat{\vecalpha}_k+\hat{\mGamma}_{k}\vecx_i)(\hat{\vecalpha}_k+\hat{\mGamma}_{k}\vecx_i)^{'}+(\hat{\vecalpha}_k+\hat{\mGamma}_{k}\vecx_i)\hat{\vecbeta}_{\eta k}^{'}+\hat{\vecbeta}_{\eta k}(\hat{\vecalpha}_k+\hat{\mGamma}_{k}\vecx_i)^{'}+E_{1ik}\hat{\vecbeta}_{\eta k}\hat{\vecbeta}_{\eta k}^{'}\Big].
\end{align*}

For the more parsimonious version of Model I, the M-step maximizes
\begin{equation*}
\ksum\lsum p_{ik}\text{logit}(\vecpi_i) =\ksum\lsum p_{ik}(\vecalpha_c + \mGamma_c \vecx_i)
\end{equation*}
with respect to the parameters $\vecalpha_c$ and $\mGamma_c$,
which may be viewed as a multinomial logistic regression with fractional observations $p_{ik}$. The updates are
\begin{align*}
\hat{\mGamma}_{c}&=\ksum\lsum p_{ik}\left(E_{5ik}\vecx_i^{'}-E_{2ik}\hat{\vecalpha}_k\vecx_i^{'}-\hat{\vecbeta}_k\vecx_i^{'}\right)\left(K\lsum\ksum p_{ik} E_{2ik} \vecx_i\vecx_i^{'}\right)^{-1},\\
\hat{\vecalpha}_c&=\frac{\ksum n_k \hat{\vecalpha}_k}{n}.
\end{align*}

\subsection{EM algorithm for Model II} %(i.e., GHD-GMM under $\vecbeta_{\eta k} = \mathbf{0}$)}

The EM algorithm for Model II was employed for parameter estimation in an analogous fashion to the algorithm for Model I described in Section~\ref{sec:modelipara}. The complete-data comprise the observed outcomes $\vecy_i$ and covariates $\vecx_i$, the class membership labels $c_{ik}$, the latent factors $\veceta_i$, and the latent $w_{ik}$, for $i=1,\ldots,n$ and $k=1,\ldots,K$. Therefore, the complete-data log-likelihood is 
\begin{align*}
l_c(\varthet) = \lsum\ksum c_{ik}[&\tlog\pi_{ik}+\tlog\phi(\vecy_i\mid\mLambda_y\veceta_i + w_{ik}\vecbeta_{yk},w_{ik}\vecTheta_k)\\
&+\tlog\phi(\veceta_i\mid\vecalpha_k+\mGamma_{k}\vecx_i,w_{ik}\noisev_k)+\tlog h(w_{ik}\mid\omega_k,\lambda_k)].
\end{align*}
 The E-step requires the computation of the conditional expectations regarding the latent factors $\veceta_i$ and the latent variables $W_{ik}$. Under this formulation, 
\[ \veceta_i \mid \vecy_i, \vecx_i, w_{ik}, c_{ik}=1\sim \mathcal{N}(\vecV_k(\inv{\noisev_k}( \vecalpha_k + \mGamma_k\vecx_i) +\mLambda_y^{'}\inv{\vecTheta_k}(\vecy_i -  w_{ik}\vecbeta_{yk}), w_{ik}\vecV_k),\]
and the conditional distribution of latent variable $W_{ik}$ given $\vecy_i, \vecx_i$, and  $c_{ik}=1$ is given by
\[
W_{ik}\mid\vecy_i,\vecx_i,c_{ik}=1\sim \text{GIG}(\psi_k^\star,\chi_{ik},\tilde{\lambda}_{k}),
\] 
with
$\psi_k^\star = \omega_k +\vecbeta_{yk}^{'}\inv\matsig_k\vecbeta_{yk}$,
$\chi_{ik} = \omega_k +\delta (\vecy_i,\vecmu_k\mid\matsig_k)$,
$\tilde{\lambda}_{k} =\lambda_k-T/2$,
where $\vecmu_k = \mLambda_y(\vecalpha_k + \mGamma_k\vecx_i)$, and $\matsig_k = \mLambda_y\noisev_k\mLambda_y^{'}+\vecTheta_k$.
Therefore, we have convenient forms for the following conditional expectations:
\begin{align*}
E_{1ik}^\star&\colonequals\Ew=\sqrt{\frac{\chi_{ik}}{\psi_k^\star}}\frac{K_{\tilde{\lambda}_k+1}(\sqrt{\psi_k^\star\chi_{ik}})}{K_{\tilde{\lambda}_k}(\sqrt{\psi_k^\star\chi_{ik}})},\\
\vspace{0.1in}
E_{2ik}^\star&\colonequals\Ewinv=\sqrt{\frac{\psi_k^\star}{\chi_{ik}}}\frac{K_{\tilde{\lambda}_k+1}(\sqrt{\psi_k^\star\chi_{ik}})}{K_{\tilde{\lambda}_k}(\sqrt{\psi_k^\star\chi_{ik}})}-\frac{2\tilde{\lambda}_k}{\chi_{ik}},\\
\vspace{0.1in}
E_{3ik}^\star&\colonequals\Elogw=\log\left(\sqrt{\frac{\chi_{ik}}{\psi_k^\star}}\right)+\frac{1}{K_{\tilde{\lambda}_k}(\sqrt{\psi_k^\star\chi_{ik}})}\frac{\partial}{\partial \tilde{\lambda}_k}K_{\tilde{\lambda}_k}(\sqrt{\psi_k^\star\chi_{ik}}),\\
\vspace{0.1in}
E_{4ik}^\star&\colonequals\Eeta=\vecV_k(\inv{\noisev_k}(\vecalpha_k+\mGamma_{k}\vecx_i)+\mLambda_y^{'}\inv{\vecTheta_k}(\vecy_i-E_{1ik}^\star\vecbeta_{ yk})),\\
\vspace{0.1in}
E_{5ik}^\star&\colonequals\Eetawinv=\vecV_k(E_{2ik}^\star(\inv{\noisev_k}(\vecalpha_k+\mGamma_{k}\vecx_i)+\mLambda_y^{'}\inv{\vecTheta_k}\vecy_i)-\mLambda_y^{'}\inv{\vecTheta_k}\vecbeta_{yk}),\\
\vspace{0.1in}
E_{6ik}^\star&\colonequals\Eetaetawinv=\vecV_k-\vecV_k(\inv{\noisev_k}(\vecalpha_k+\mGamma_{k}\vecx_i)+\mLambda_y^{'}\inv{\vecTheta_k}\vecy_i)\vecbeta_{yk}^{'}\inv{\vecTheta_k}\mLambda_y\vecV_k\\
&+E_{2ik}^\star\vecV_k(\inv{\noisev_k}(\vecalpha_k+\mGamma_{k}\vecx_i)+\mLambda_y^{'}\inv{\vecTheta_k}\vecy_i)(\inv{\noisev_k}(\vecalpha_k+\mGamma_{k}\vecx_i)+\mLambda_y^{'}\inv{\vecTheta_k}\vecy_i)^{'}\vecV_k,\\
&-\vecV_k\mLambda_y^{'}\inv{\vecTheta_k}\vecbeta_{yk}^{'}(\inv{\noisev_k}(\vecalpha_k+\mGamma_{k}\vecx_i)+\mLambda_y^{'}\inv{\vecTheta_k}\vecy_i)^{'}\vecV_k+E_{1ik}^\star\vecV_k\mLambda_y^{'}\inv{\vecTheta_k}\vecbeta_{yk}\vecbeta_{yk}^{'}\inv{\vecTheta_k}\mLambda_y\vecV_k.
\end{align*}

At each E-step, the values of $E_{1ik}^\star,E_{2ik}^\star,\ldots,E_{6ik}^\star$ are updated. We also update the value of the class membership variable $c_{ik}$ using
\begin{equation*}
p_{ik}^\star\colonequals%\mathbb{E}\left[c_{ik} \mid \vecy_i,\vecx_i\right]=
\frac{\pi_{ik} f_{\text{GHD}_T}(\vecy_i ; \tilde{\lambda}_k,\omega_k,\vecmu_k,\matsig_k,\vecbeta_{yk})}{\sum_{l=1}^{K} \pi_{il} f_{\text{GHD}_T}(\vecy_i ; \tilde{\lambda}_l,\omega_l,\vecmu_l,\matsig_l,\vecbeta_{yl})}.
\end{equation*}
At each M-step, the following model parameters are obtained by maximizing the conditional expected value of $l_c(\varthet)$ and are updated sequentially. The updates for $\vecpi_{ik}$, $\vecalpha_c$, $\mGamma_c$, $\tilde{\lambda}_k$, and $\omega_k$ are similar to those used in Appendix~\ref{sec:modelipara}. We update the skewness parameter $\vecbeta_{yk}$ using
\[
	\hat{\vecbeta}_{yk} = \frac{\lsum p_{ik}^\star (\vecy_i - \mLambda_yE_{4ik}^\star)}{\lsum p_{ik}^\star E_{1ik}^\star}
\]
and the measurement error $\vecTheta_k$ using
\begin{align*}
	\hat{\vecTheta}_k = \frac{1}{n_k} \lsum p_{ik}^\star \Big(&E_{2ik}^\star\vecy_i\vecy_i^{'}-\vecy_iE_{5ik}^{\star'}\mLambda_y^{'}-\vecy_i\hat{\vecbeta}_{yk}^{'}-\mLambda_yE_{5ik}^\star\vecy_i^{'}+\mLambda_yE_{6ik}^\star\mLambda_y^{'}+\mLambda_yE_{4ik}^\star\hat{\vecbeta}_{yk}^{'}\\
	&-\hat{\vecbeta}_{yk}\vecy_i^{'}+\hat{\vecbeta}_{yk}E_{4ik}^{\star'}\mLambda_y^{'}+E_{1ik}^\star\hat{\vecbeta}_{yk}\hat{\vecbeta}_{yk}^{'}\Big),
\end{align*}
where $n_k = \lsum p_{ik}^\star$. We update $\mGamma_k$, $\vecalpha_k$, and $\noisev_k$ sequentially using
\begin{align*}
	\hat{\mGamma}_k &=  \left[\lsum p_{ik}^\star (E_{5ik}^\star-E_{2ik}^\star\hat{\vecalpha}_k)\vecx_i^{'}\right]\left[\lsum p_{ik}^\star E_{2ik}^\star\vecx_i\vecx_i^{'}\right]^{-1},\\
	\hat{\vecalpha_k} &= \frac{\lsum p_{ik}^\star (E_{5ik}^\star-E_{2ik}^\star\hat{\mGamma}_k\vecx_i)}{\lsum p_{ik}^\star E_{2ik}^\star},\\
	\hat{\noisev}_k &= \frac{1}{n_k} \lsum p_{ik}^\star \left[E_{6ik}^\star - E_{5ik}^\star(\hat{\vecalpha}_k + \hat{\mGamma}_{k}\vecx_i)^{'} - (\hat{\vecalpha}_k + \hat{\mGamma}_{k}\vecx_i)E_{5ik}^{\star'} +E_{2ik}^\star(\hat{\vecalpha}_k +\hat{\mGamma}_{k}\vecx_i)(\hat{\vecalpha}_k + \hat{\mGamma}_{k}\vecx_i)^{'}\right].
\end{align*}

\subsection{EM algorithm for Model III} %(i.e., GST-GMM under $\vecbeta_{yk} = \mathbf{0}$)}
Similarly, parameter estimation for Model III is carried out within the EM algorithm framework. Suppose we observe the outcome $\vecy_i$ and the covariates $\vecx_i$ from a GMM with skew-t random effects as in \eqref{eqn:gstgmm} but with $\vecbeta_{yk}=\mathbf{0}$. There are three sources of unobserved data: the latent categorical variables $\vecc_i$, the latent growth factors $\veceta_i$, and the latent $w_{ik}$. The complete-data log-likelihood can be expressed as follows:
\begin{align*}
\mathcal{L}_c(\varthet) = \lsum\ksum \pi_{ik}\big[&\tlog\pi_{ik} +\tlog \phi(\vecy_i\mid\mLambda_y\veceta_i,w_{ik}\vecTheta_k)\\
&+\tlog\phi(\veceta_i\mid\vecalpha_k+\mGamma_{k}\vecx_i+w_{ik}\vecbeta_{\eta k},w_{ik}\noisev_k)+\tlog f(w_{ik}\mid \nu_k/2,\nu_k/2)\big],
\end{align*}
where %$\pi_{ik}$ is defined as above in \eqref{eqn:cx} and 
$f(w_{ik}\mid \nu_k/2,\nu_k/2)$ is the density of the inverse Gamma distribution.

The E-step requires the computation of the expected value of the complete-data log-likelihood. Note that $W_{ik}\mid\vecy_i,\vecx_i,c_{ik}=1\sim \text{GIG}(\psi_k^{\ast},\chi_{ik}^{\ast},\lambda_{k}^{\ast})$ with $\psi_k ^{\ast}= \vecbeta_{\eta k}^{'}\mLambda_y^{'}\inv\matsig_k\mLambda_y\vecbeta_{\eta k}$, $\chi_{ik}^{\ast} = \nu_k +\delta (\vecy_i,\vecmu_k\mid\matsig_k)$, and $\lambda_{k}^{\ast} = -{(\lambda_k+T)}/{2}$. Therefore, we have convenient forms for the following expected values:
\begin{align*}
E_{1ik}^{\ast}&\colonequals\Ew=\sqrt{\frac{\chi_{ik}^{\ast}}{\psi_k^{\ast}}}\frac{K_{\lambda_k^{\ast}+1}(\sqrt{\psi_{k}^{\ast}\chi_{ik}^{\ast}})}{K_{\lambda_k^{\ast}}(\sqrt{\psi_{k}^{\ast}\chi_{ik}^{\ast}})},\\
\vspace{0.1in}
E_{2ik}^{\ast}&\colonequals\Ewinv=\sqrt{\frac{\psi_k^{\ast}}{\chi_{ik}^{\ast}}}\frac{K_{\lambda_k^{\ast}+1}(\sqrt{\psi_{k}^{\ast}\chi_{ik}^{\ast}})}{K_{\lambda_k^{\ast}}(\sqrt{\psi_{k}^{\ast}\chi_{ik}^{\ast}})}-\frac{2\lambda_k^{\ast}}{\chi_{ik}^{\ast}},\\
\vspace{0.1in}
E_{3ik}^{\ast}&\colonequals\Elogw=\log\left(\sqrt{\frac{\chi_{ik}^{\ast}}{\psi_k^{\ast}}}\right)+\frac{1}{K_{\lambda_k^{\ast}}(\sqrt{\psi_{k}^{\ast}\chi_{ik}^{\ast}})}\frac{\partial}{\partial \lambda_k^{\ast}}K_{\lambda_k^{\ast}}(\sqrt{\psi_{k}^{\ast}\chi_{ik}^{\ast}}).
\vspace{0.1in}
\end{align*}
We also need the expected value of the class membership, i.e.,
\begin{equation*}
\tau_{ik}\colonequals\mathbb{E}\left[C_{ik} \mid \vecy_i,\vecx_i\right]=\frac{\pi_{ik} f_{\text{GHD}_T}(\vecmu_k,\matsig_k,\mLambda_y\vecbeta_{\eta k},v_k)}{\sum_{l=1}^{K} \pi_{il} f_{\text{GHD}_T}(\vecmu_l,\matsig_l,\mLambda_y\vecbeta_{\eta l},v_l)},
\end{equation*}
as well as the following conditional expectations, which are similar to those derived in the E-step of parameter estimation for the GHD-GMM:
\begin{align*}
E_{4ik}^{\ast}&\colonequals\Eeta=\vecV_k(\noisev_k^{-1}(\vecalpha_k+\mGamma_{k}\vecx_i+E_{1ik}^{\ast}\vecbeta_{\eta k})+\mLambda_y^{'}\vecTheta_k^{-1}\vecy_i),\\
E_{5ik}^{\ast}&\colonequals\Eetawinv=E_{2ik}^{\ast}\vecV_k(\noisev_k^{-1}(\vecalpha_k+\mGamma_{k}\vecx_i)+\mLambda_y^{'}\vecTheta_k^{-1}\vecy_i)+\vecV_k\noisev_k^{-1}\vecbeta_{\eta k},\\
E_{6ik}^{\ast}&\colonequals\Eetaetawinv=\vecV_k+\vecV_k(\noisev_k^{-1}(\vecalpha_k+\mGamma_{k}\vecx_i)+\mLambda_y^{'}\vecTheta_k^{-1}\vecy_i)\vecbeta_{\eta k}\noisev_k^{-1}\vecV_k\\
&+E_{2ik}^{\ast}\vecV_k(\noisev_k^{-1}(\vecalpha_k+\mGamma_{k}\vecx_i)+\mLambda_y^{'}\vecTheta_k^{-1}\vecy_i)(\noisev_k^{-1}(\vecalpha_k+\mGamma_{k}\vecx_i)+\mLambda_y^{'}\vecTheta_k^{-1}\vecy_i)^{'}\vecV_k,\\
&+\vecV_k\noisev_k^{-1}\vecbeta_{\eta k}^{'}(\noisev_k^{-1}(\vecalpha_k+\mGamma_{k}\vecx_i)+\mLambda_y^{'}\vecTheta_k^{-1}\vecy_i)^{'}\vecV_k+E_{1ik}^{\ast}\vecV_k\noisev_k^{-1}\vecbeta_{\eta k}\vecbeta_{\eta k}^{'}\noisev_k^{-1}\vecV_k.
\end{align*}
The M-step requires the computation of the parameter updates to maximize the conditional expected value of the complete-data log-likelihood. In this step, the parameter updates for $\vecalpha_c, \mGamma_c, \vecalpha_k,\vecbeta_{\eta k}, \vecTheta_k, \noisev_k$, and $\mGamma_k$ are obtained in closed form and are similar to those derived in the M-step of parameter estimation for the GHD-GMM and, hence, are omitted here. To obtain the update for $\nu_k$, we solve the equation
\begin{equation}
\tlog\left(\frac{\nu_k}{2}\right) + 1 - \varphi\left(\frac{\nu_k}{2}\right) - \frac{1}{ n_k} \lsum \tau_{ik} \left(E_{3ik}^{\ast} + E_{2ik}^{\ast}\right) = 0
\end{equation}
for $\nu_k$, numerically, where $n_k=\lsum \tau_{ik}$.

\end{document}